\documentclass[reprint,amssymb, amsmath, aps, prb]{revtex4-2}

\usepackage{bm}%
\usepackage[colorlinks=true,linkcolor=blue]{hyperref}
\expandafter\ifx\csname package@font\endcsname\relax\else
 \expandafter\expandafter
 \expandafter\usepackage
 \expandafter\expandafter
 \expandafter{\csname package@font\endcsname}%
\fi
\usepackage[justification=justified]{caption}
\usepackage{multirow}
\usepackage{etoolbox}
\usepackage[english]{babel}

\BeforeBeginEnvironment{figure}{\vskip 0ex}
\AfterEndEnvironment{figure}{\vskip 0ex}

\usepackage[justification=justified]{subcaption}
\usepackage{graphicx}
\usepackage[dvipsnames]{xcolor}
\usepackage{tcolorbox}
\tcbuselibrary{breakable}

\begin{document}

\title{Oxygen deficiency and migration mediated electric polarization in Fe,Co-substituted SrTiO$_{3-\delta}$}

\author{E. A. Cort\'es$^{1}$, S. P. Ong$^2$, C. A. Ross$^3$, and J. M. Florez$^{1,3}$}%
\email{juanmanuel.florez@usm.cl, jmflorez@mit.edu}
\affiliation{$^1$Grupo de Simulaciones, Departamento de F\'isica, Universidad T\'ecnica Federico Santa Mar\'ia, Casilla 110-V, Valpara\'iso, Chile\\
$^2$Department of NanoEngineering, University of California, San Diego,
9500 Gilman Drive, La Jolla, California 92093, USA\\
$^3$Department of Materials Science and Engineering, Massachusetts Institute of Technology, 77 Massachusetts Avenue, Cambridge, Massachusetts 02139, USA}%


\begin{abstract}
We use density functional theory (DFT) calculations to show that oxygen vacancies ($v_\mathrm{O}$) and mobility induce noncentrosymmetric polar structures in SrTi$_{1-x-y}$Fe$_{x}$Co$_{y}$O$_{3-\delta}$ ($x=y=0.125$) with $\delta = \{0.125, 0.25\}$, enhance the saturation magnetization and give rise to large changes in the electric polarization $\vert\Delta P\vert$. We present an intuitive set of rules for SrTiFeCoO$_{3-\delta}$ (STFC), which are based on the interplay between (Co/Fe)-$v_\mathrm{O}$ defects, magnetic cations coordination and topological vacancy disorder. STFC structures convey layered crystals with sheets of linear organized O$_{4,5,6}$-coordinated Fe-Co pairs, sandwiched with layers of O$_{5}$-coordinated Ti.
Co,Fe-$v_\mathrm{O}$ defects are the source of the crystal distortions, cations off-centering and bending of the oxygen octahedra, which added to the charge redistribution mediated by $v_\mathrm{O}$, the cations electronegativity and valence states trigger an effective electric polarization.
Oxygen migrations for $\delta=0.125$ provides us with $\vert\Delta \mathbf{P}\vert$ $>\sim10 \mu$C/cm$^2$ due to a quantum-of-polarization differences between $\delta=0.125$ structures. Increasing the deficiency to $\delta=0.25$ yields $\vert\Delta \mathbf{P}\vert$ whose O-migration resolved polarization for $\delta=0.25$ is $>\sim3 \mu$C/cm$^2$ in the worst case scenario. Magnetism is dominated by the Fe,Co spin states for $\delta=0.125$ while there is a raid of Ti magnetic moments ($\sim1\mu_{B}$) for  $\delta=0.25$. Magnetic and electric order parameters change for variations of $\delta$ or oxygen migrations for a given deficiency. Our results capture characteristics observed in the end-members of the series SrTi(Co,Fe)O$_{3}$, and suggest the existence of a broader set of rules for oxygen deficient multiferroic oxides.
\end{abstract}
\maketitle

\section{Introduction}
Materials that possess simultaneously at least two ferroic orders (ferroelectricity, ferroelasticity, and ferro/ferri/antiferromagnetism) are described as multiferroic \cite{MultiferroicsSpaldin2019}. The search for such materials has expanded to several classes of systems, of which the magnetic perovskites provide outstanding examples with a wide range of cation compositions and oxygen stoichiometries  \cite{multiferroico1,MnPerovskite_interplay,MultiferroicsSpaldin2019}. 
Among transition metal (TM) perovskites, SrTi$_{1-x}$Fe$_x$O$_{3-\delta}$ (STF) and SrTi$_{1-x}$Co$_{x}$O$_{3-\delta}$ (STC) both display magnetization that depends on their oxygen content, with typically higher  magnetization at higher levels of oxygen deficiency ($\delta$) \cite{OdefSTF, STC, STC2018}. On the other hand, at room-temperature, stoichiometric SrTiO$_{3}$ (STO) is a nonmagnetic paraelectric. At low temperatures it presents an antiferrodistortive structural change that suppresses the ferroelectric (FE) ordering, as well as quantum fluctuations that forbid the FE phase-transition \cite{STO_Review,STO_comp_AFD-FE}. 
Several mechanisms promote a FE phase in STO, e.g. the application of electric fields, A-site cation substitution, strain in thin films \cite{STO_Efield,STO_RTFE_strain} and non-stoichiometric solid solutions have all been proposed to lead to a ferroelectric response \cite{STO_strainfree_FE,STO_RTFE_tetra}. Defects such as coupled Sr ($v_{Sr}$) and O ($v_\mathrm{O}$) vacancies, $v_{Sr}$ with interstitial I$_{\mathrm{Ti}}$, anti-site Ti$_{\mathrm{Sr}}$ and Sr$_{\mathrm{Ti}}$ and their coupling with $v_\mathrm{O}$\cite{STO_SrOO_vac,STO_Ti_antisite,STO_Pol_defects} have been suggested to promote ferroelectricity in STO with  Sr/Ti atomic ratio close to 1 \cite{STO_FE_SrTi_ratio}. A-site defects such as $v_{\mathrm{Sr}}$ can also mediate polar effects in (111)-oriented and (001)-bulk--terminated perovskites based on SrTiO$_{3}$\cite{STO111,STO001}.

Oxygen deficiency has been used to enhance multiferroism in PbTiO$_{3}$ \cite{ferro1} and ferroelectric switching in Si-doped HfO$_{2}$ \cite{ferro2}. Ferroic parameters in YMnO$_{3}$ \cite{ferro3} and strained SrMnO$_{3}$ \cite{ferro4} could be manipulated using $\delta$ as well as the ferroelectric response in perovskite structured relaxors \cite{ferro5}. Multiferroism in oxygen deficient SrFeO$_{3}$ nanoparticles was also studied \cite{ferro6}, and the magnetic ordering followed the tendencies of the $v_\mathrm{O}$ modulated magnetism in STF \cite{OdefSTF} while the FE parameter was not clearly associated to the $v_\mathrm{O}$. STF also displays ferroelectricity in nanocrystalline thin-films form with a saturation polarization up to $6 \mu$C/cm$^2$ depending on the Fe concentration \cite{STF_multiferroic}. Also, a comparable field-driven polarization of up to $\sim{1}$ $\mu$C/cm$^2$ was realized in Fe-doped Ti-rich STO multiferroics at room temperature \cite{STF_multiferroic2}. In the case of STC,  although its FE properties have not been reported, preliminary work \cite{APS} suggests that STC could share some FE characteristics with O-deficient STO. STC presents a relatively large band gap and room temperature magnetization \cite{STC_ovac,STC_ovac_exp} as well as  off-centering features induced by incorporation of Co \cite{Sluchinskaya:2019hu} or through the Co-$v_\mathrm{O}$-Ti distortions \cite{STFC_Ox_hybrid}, which are ingredients of the STO-based ferroism discussed in this work.\\

The use of oxygen deficiency as a tool to tailor ferroic order parameters requires a profound knowledge of the roles of the defects density and vacancies distribution, ABO$_{3}$ cations symmetry and ratio as well as of the TM electronic features such as valence states, radii, electronegativity, stabilizing hybridizations and occupancies \cite{OdefSTF,STC,STFC_Ox_hybrid}. Although physical/chemical synthesis are closing to more precise methods for controlling several of the aforementioned factors \cite{OdefSTF,STFC_Ox_hybrid}, to obtain specific polarization/magnetization values desperately requires theoretical/simulational insights that can help to both narrow the large space of configurations conformed by the stoichiometries and A/B crystalline solutions (S6) for practical applications, and understand the microscopic mechanisms that underlie the origin of such ferroic orders in oxygen deficient perovskites. In this work, we theoretically explore the perovskite SrTi$_{1-x-y}$Fe$_{x}$Co$_{y}$O$_{3-\delta}$ (STFC) \cite{STFC_Ox_hybrid} from the three specific scopes mentioned above i.e., using different ${\delta}$ values while considering several $v_\mathrm{O}$ distributions for each ${\delta}$, substituting B cations with Fe and Co in distinct crystalline configurations, and modulating the TM spin states among the possible valance solutions and polarizations. STFC-based systems have been experimentally studied in the context of oxygen transport in membranes \cite{STFC_Ox_memb}, electrical conductivity in STO-based energy applications and memristor \cite{stobased_applications}, as well as of oxygen electrodes for solid oxide cells \cite{stobased_applications,solidcellsstfcbased}. The ferroelectric features of STFC have not been addressed, though there are indications that this material versatility could extend to multiferroism as we shall show here.\\

In Fe substituted STO the saturation magnetization can be modulated as well as ferroelectric features are induced for low ${x}$ values (Fe$_{x}$, ${y=0}$)\cite{OdefSTF,STF_multiferroic2}. STF tends to stabilize antiferromagnetic solutions for low and high ends ${\delta}$ values while Co substituted STO is predominantly ferromagnetic for low deficiencies and not diluted ${y}$ substitutions (Co$_{y}$, ${x=0}$)\cite{OdefSTF,STC,STC2018,STF_molecular}. STC band-gap increases with ${\delta}$, with presumably small but finite electric polarizations\cite{STC,APS}; STFC presents an oxygen deficiency modulated band-gap that maximizes at intermediate ${\delta}$ along with the saturation magnetization\cite{STFC_Ox_hybrid}. On the other hand, SrFe$_{1-z}$Co$_{z}$O$_{3-\delta}$, without the magnetic dissolvent and insulating role of Ti, is able to display voltage tuned magnetic response \cite{SCF_voltage}, which for low end ${z}$ values nanostructured systems also display multiferroism \cite{ferro6}. All these factors suggest that Fe/Co substituted STO could combine these aforementioned order parameters within a realizable range of oxygen deficiency, moreover, STFC could help to understand better the role of ${\delta}$ in the triggering of multiferroic parameters i.e., mechanisms known so far within proper or/and improper ferroelectricity frameworks \cite{multiferroico1,MnPerovskite_interplay,MultiferroicsSpaldin2019} do not seem to completely account for several features of oxygen deficient ABO$_{3-\delta}$, therefore, computational engineering of nonstoichiometric oxygen solutions could show us one path toward synthesizing room-temperature multiferroics, given that we can figure how such new mechanisms work, beside innovating on how to manage the oxygen content in benefit of them.\\

In this work, we primordially focused on the ferroelectric response of oxygen deficient STFC. We demonstrate by using density functional theory (DFT) calculations that oxygen vacancies ($v_\mathrm{O}$)  in SrTi$_{0.75}$Fe$_{0.125}$Co$_{0.125}$O$_{3-\delta}$ with  $\delta = \{0.125, 0.25\}$ induce non-centrosymmetric polar structures, which support large electric polarization changes $\vert\Delta \mathbf{P}\vert$ through $v_\mathrm{O}$ migration and/or O-deficiency changes that are compatible with a robust magnetization. 
We explore a variety of $v_\mathrm{O}$ configurations and show how $\vert\Delta \mathbf{P}\vert$ varies within $\sim{23\mu}$C/cm$^2$ for lowest-energy stabilized defected structures, with $\delta = 0.125$ systems displaying the highest resolved polarization differences. Our results capture key aspects of recent results on Fe-Co substituted STO, and give a further insight into the electronic and structural mechanisms that would enable the realization of multiferroics by using specific defects and/or cation distributions in ABO$_{3-\delta}$. 
\section{DFT modeling}
Spin-polarized DFT calculations were performed using the \textit{Vienna Ab-initio Simulation Package} (VASP 5.4) \cite{vasp3}  within the projector-augmented-wave (PAW) method\cite{PAW} and an energy cutoff of 500 eV. \textit{k}-point grids of $4\times4\times4$ for relaxation calculations as well as $6\times6\times6$ \textit{k}-points for static calculations were used. The valence electrons included in the chosen pseudopotentials for Sr, Ti, Fe, Co and O are $4s^2 4p^6 5s^2$, $3d^3 4s^1$, $3d^7 4s^1$, $3d^8 4s^1$ and $2s^2 2p^4$, respectively. 

All ions and supercell parameters were relaxed until atomic forces were below 0.05 and 0.01 eV/\AA, for large and intermediate $\delta$ values respectively. Oxygen migration barriers were computed using the nudged elastic band (NEB) method \cite{NEB}, with forces converged below 100 meV/$\text{\AA}$ and energies below $10^{-5}$ eV. The generalized gradient approximation (GGA) within the Dudarev approach for the $+U$ corrected Hubbard model (GGA+U) was used for $d$ electrons of the TM \cite{ldau_vasp}. 
\begin{figure*}[t]
\centering
\includegraphics[width=1.4\columnwidth]{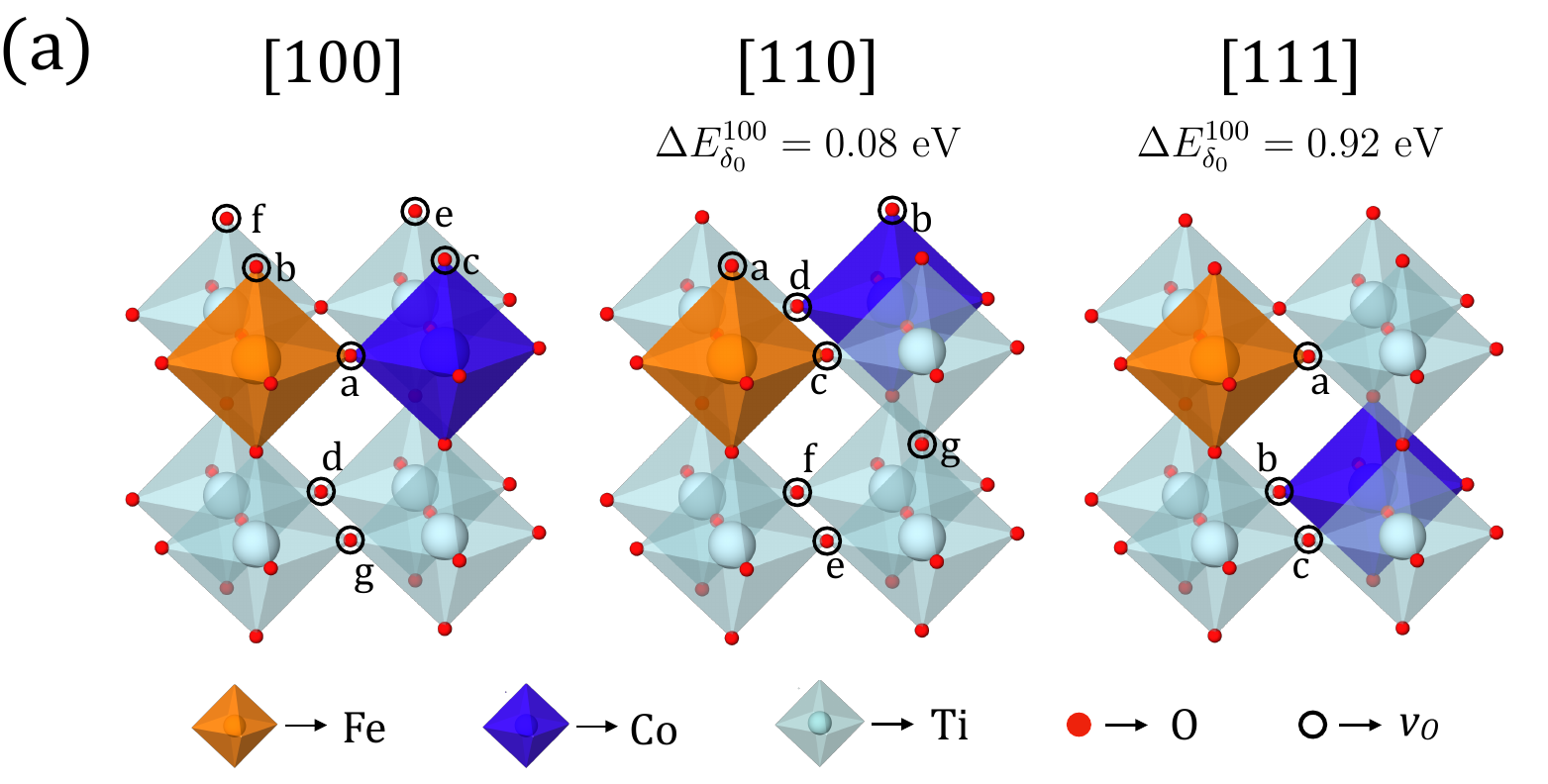}
\includegraphics[width=1.4\columnwidth]{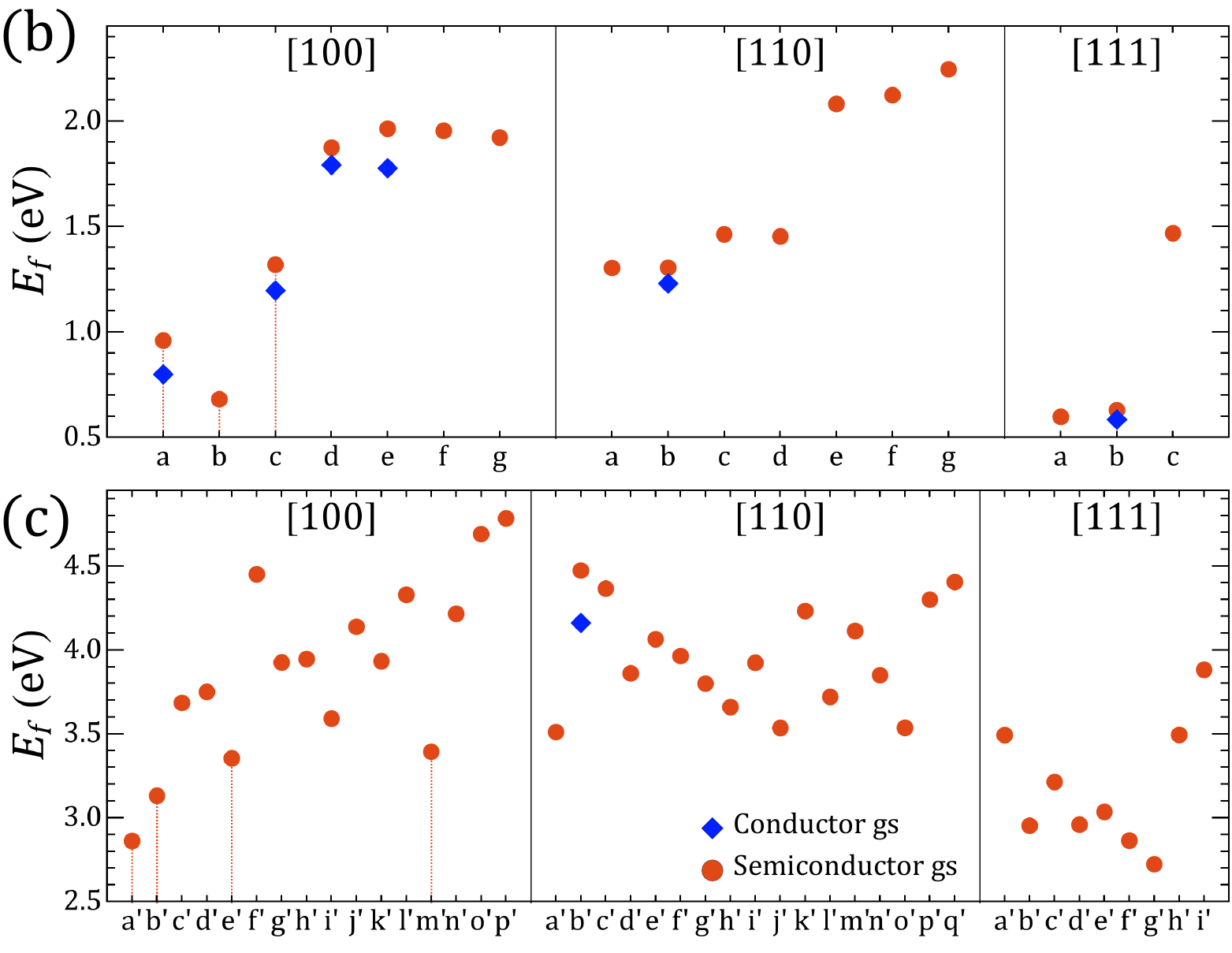}
\caption{(a) Fe-Co arrangements in a $2\times2\times2$ perovskite cell and $v_\mathrm{O}$ positions for $\delta_{01}$. ${\Delta}E_{\delta_{0}}^{100}=$ difference with [100] configuration.
(b) Formation energies $E_f$ for the gs (ground state) found for all $v_\mathrm{O}$ in the three Fe-Co configurations for $\delta_{01}$, and (c) $E_f$ for $\delta_{02}$; $v_\mathrm{O}$ positions are labeled with letters as defined in Figure S1 of the Supplemental Material\cite{Supplemental}. Solutions marked with vertical lines, $a$, $b$, $c$ ($\delta_{01}$) and $a'$, $b'$, $e'$ and $m'$ ($\delta_{02}$), are used for subsequent analysis.}
\label{fig:1V_FE}
\end{figure*}
The $U$ values used here for Fe, Co and Ti are 3, 9 and 8 eV, respectively, and were chosen based on an extensive search for lattice parameters, local magnetic moments and band gaps behaviors that would best represent end members SrTiO$_3$, SrFeO$_3$ and SrCoO$_3$, as well as the intermediate solutions SrTi$_{1-x}$Fe$_{x}$O$_3$ and SrTi$_{1-x}$Co$_{x}$O$_3$ with $x= \{0.125,0.25\}$ \cite{OdefSTF, STC} (S2,S3); such U values also capture distinctive features concluded from hybrid calculations for STFC \cite{STFC_Ox_hybrid}, and are used here to also calculate the oxygen migration barriers and paths.

The electric polarization was calculated by using the Berry's Phase approach \cite{ModPol4}. In this work we do not consider ferroelectric response originated in metallic systems\cite{metallicferro1,metallicferro2}, for which an intuitive formation-energy screening of the defected structures is used to narrow the configurations to the more interesting insulating $v_\mathrm{O}$ states. The core of the calculations using such a Berry's Phase theory was done with the implementation provided by VASP\cite{ModPol4}, and the quantum-of-polarization space analysis, which is discussed in section IV, was performed with homemade software. The pre- and post-processing of data required for this work were performed with VESTA \cite{VESTA}, CrystalMaker\textsuperscript{\textregistered} \cite{crystalMaker} and pymatgen-based home programs \cite{pymatgen}.\\

SrTi$_{0.75}$Fe$_{0.125}$Co$_{0.125}$O$_{3-\delta}$ prototypes are modeled using $2\times2\times2$ supercells, with the pair of Fe-Co ions aligned along [100], [110], and [111] directions as shown in upper panel Figure \ref{fig:1V_FE}(a). 
Studies of order and coordination of TM-cations have recently given new insights into the stabilization of multiferroic parameters \cite{OdefSTF,STC,STFC_Ox_hybrid,ordercatios-1,ordercatios-2}, the anionic-redox paradigm and the voltage decay in Li-ion batteries \cite{Assat:2018hx,Myeong:2018gd}. The breaking of atomic ordering was used to faster oxygen diffusion in deficient strained oxides \cite{breakingoxydifucion2021} as well as block structured crystals rather diluted solid-solutions were unveiled using atomic resolution X-ray emission in niobium-tungsten complex oxides\cite{chemicalorderin2021nature}, pointing to the importance of the ordered local character in these materials.

Once the three configurations above are analyzed, which is done by testing all the valence states admissible for the TM stoichiometry and the possible spin polarizations, $v_\mathrm{O}$ are created by removing one ($\delta = 0.125 \Rightarrow \delta_{01}$) or two ($\delta = 0.25 \Rightarrow \delta_{02}$) oxygen atoms out of the relaxed $\delta={0} \Rightarrow \delta_{0}$ systems, and the resulting structures are grouped following a symmetry analysis of the configuration space with a tolerance of between 10$^{-3}$-10$^{-4}$\AA (S6). For $\delta_{01}$, such an analysis leads us to 7 nonequivalent O-deficient configurations for [100] and [110] TM alignments. The [111] alignment has 3 configurations, all of them shown in Figure \ref{fig:1V_FE}(a). In the case of $\delta_{02}$, there exist a total of 156 nonequivalent double-vacancy structures. According to each configuration, we have again considered all the atomic valence states, which are reflected in several possibles high and low Pauli states for the TM, as well as the possible combinations for the FM or AFM Co-Fe exchange coupling. The valences of the cations are such that we maintain a neutral supercell for a given oxygen deficiency, and charged defects are not considered because our calculations suggest spontaneous charging of the $v_\mathrm{O}$ is negligible, even for an excess or deficiency of electrons, in agreement with hybrid simulations \cite{STFC_Ox_hybrid}, therefore, a charge-background is not determinant from our scope\cite{vaspbackground}. We initialized the systems with those combinations, then relaxations without magnetic constraints are performed\cite{STC}.\\

 \begin{table*}[ht]
    \centering
    \caption{Semiconductor ground states (gs): $[100]_{a,b,c}$ for $\delta_{01}$, Figure \ref{fig:1V_FE}(a); and  $[100]_{a',b',e',m'}$ for $\delta_{02}$, Figure \ref{fig:1V_FE}(b). $\Delta E$ with respect to the lowest gs. Space group was obtained using a tolerance of $0.05$ \AA{}.}
    \label{tab:1V2V_SmallTable}
    \begin{ruledtabular}
        \begin{tabular}{lccccccccccccc}
        & $a$ (\AA) & $b$ (\AA) & $c$ (\AA) & Volume (\AA\textsuperscript{3}) & $\mu_B/\text{Fe}$ & $\mu_B/\text{Co}$ & $\mu_B/u.c.$ & Order & $\Delta E$ (meV/$f.u.$) & $E_{gap}$ (eV)  & Space group \\
        $[100]_a$ & 8.01113 & 8.00104 & 8.00101 & 512.843 & 3.96 & 3.21 & 0.87 & \text{AFM} & 17.1 & 0.0 & P4mm\\
        $[100]_a$ & 8.03717 & 7.9533 & 8.01124 & 512.094 & 4.10 & 3.43 & 8.90 & \text{FM} & 37.1 & 0.39 & Pmm2 \\
        $[100]_b$  & 8.03101 & 7.97895 & 8.03229 & 514.696 & 3.37 & 2.90 & 0.84 & \text{AFM} & 0.8 & 0.40 & P1 \\
        $[100]_b$  & 8.03693 & 7.97111 & 8.03031 & 514.442 & 3.41 & 2.91 & 6.87 & \text{FM} & 0. & 0.45 & P1 \\
        $[100]_c$  & 8.00861 & 7.99924 & 8.03576 & 514.793 & 3.60 & 2.87 & 6.91 & \text{FM} & 66.8 & 0.0 & Pm \\
        $[100]_c$  & 7.9939 & 8.01455 & 8.02041 & 513.847 & 4.12 & 3.32 & 0.91 & \text{AFM} & 82.1 & 0.64 & Pmm2 \\
        \hline
        $[100]_{a\text{\textquoteright}}$ & 7.93281 & 8.0405 & 8.04098 & 512.883 & 3.63 & 2.88 & 1.00 & \text{AFM} & 0. & 1.20 & P4/mmm \\
        $[100]_{a\text{\textquoteright}}$ & 7.92617 & 8.04598 & 8.04597 & 513.122 & 3.64 & 2.87 & 7.02 & \text{FM} & 20.2 & 1.14 & P4/mmm \\
        $[100]_{b\text{\textquoteright}}$ & 8.08034 & 8.07529 & 7.97517 & 520.313 & 4.08 & 2.91 & 6.88 & \text{FM} & 33.7 & 1.56 & P1 \\
        $[100]_{b\text{\textquoteright}}$ & 8.10709 & 8.03554 & 8.00719 & 521.573 & 4.07 & 2.85 & 0.88 & \text{AFM} & 46.1 & 1.55 & Pm \\
        $[100]_{e\text{\textquoteright}}$ & 7.96727 & 8.09836 & 8.04881 & 519.323 & 4.07 & 2.87 & 0.85 & \text{AFM} & 61.6 & 1.32 & Pm \\
        $[100]_{e\text{\textquoteright}}$ & 7.96523 & 8.10383 & 8.04897 & 519.551 & 4.09 & 2.92 & 6.84 & \text{FM} & 64.8 & 1.35 & P1 \\
        $[100]_{m\text{\textquoteright}}$ & 8.03584 & 8.10503 & 7.9565 & 518.213 & 4.08 & 2.87 & 2.84 & \text{AFM} & 66.5 & 1.77 & Pmm2 \\
        $[100]_{m\text{\textquoteright}}$ & 8.04113 & 8.10629 & 7.95194 & 518.337 & 4.09 & 2.92 & 8.83 & \text{FM} & 73.1 & 1.73 & Pmm2\\
        \end{tabular}
    \end{ruledtabular}
\end{table*}
Although O-deficiency is commonly observed in experiments a path to use it to selectively enhance ferroic parameters is still unknown. The data obtained in the process described above and in what follows, allows us to have a first glance of what an statistical response could be given our perovskite stabilize into those states. A previous experimental/theoretical observation can now be applied in order to have a more representative group of states: we seek for materials with large local magnetization, which from our approach means looking for crystals that present at least one $v_\mathrm{O}$ coordinating either a Fe or Co cation\cite{Supplemental} as it favors high-spin states \cite{STC,STFC_Ox_hybrid} and therefore a larger magnetization/unit-cell. This last statement is more discussed in next sections. Now, we have narrowed our configurations space down to 59 structures which are displayed in Figure \ref{fig:1V_FE} and Figure S1. We then analyze the formation energies associated to those last configurations, remembering that although for STC and STF the $[100]$, $[110]$ and $[111]$ alignments display energetically distinct states for the same magnetic saturation \cite{OdefSTF, STC, STC2018,STFC_Ox_hybrid}, and that for most cations chemical potentials are equal to the DFT ground states energies, for some elements at T=0 the ground state could be not adequate as a reference state and corrections to the energies should be made e.g., to elements with structural phase transitions below room temperature (Ti, Na, Sn), diatomic gases (O), and elements subjected to +U corrections\cite{eformationdata}. Moreover, different numbers of coordination due to vacancies distributions are related to the ions radii, which in turn can be related to various valance states, therefore, comparing Equation \ref{formation} results would be more suitable than total energy comparisons. The expression for the energy formation $E_f$ that have evaluated is such that: 
\begin{center}
\begin{equation}
E_f = E_{\text{$v_\mathrm{O}$}} - E_{\delta_{0}} + N_{\text{$v_\mathrm{O}$}} \mu_{\text{O}}  
\label{formation}
\end{equation}
\end{center}
where $E_{\text{$v_\mathrm{O}$}}$ is the total energy of the defective crystals after relaxation, and $E_{\delta_{0}}$ is the total energy of the crystals corresponding to $\delta_{0}$.  $N_{\text{$v_\mathrm{O}$}}$ is the number of removed oxygen atoms, and $\mu_{\text{O}}$ the chemical potential with respect to $\text{O}_2$ \cite{SCO_OV}. The energies obtained with Equation \ref{formation} are also corrected by using the fitted experimental enthalpies, as GGA/PBE calculations for $\text{O}_2$ tend to suffer from overbinding errors \cite{O2CorrectionCeder,O2Correction2}. The resulting energies are presented in Figure \ref{fig:1V_FE} according to the respective orientations of the Fe-Co pairs i.e., along [100], [110], and [111] crystalline axes.\\

In what follows, we use the data in Figure \ref{fig:1V_FE} to evaluate the ferroic order parameters of STFC using a representative set of states that are selected as the next section describes. Those ferroic orders are contrasted in terms of its oxygen deficiency $\delta$ and for each $\delta$ an oxygen migration analysis connecting the lowest-energy vacancy states is performed in search of those insights that lead us to further understand how to design a multiferroic perovskite using its oxygen content.

\section{Representative configurations}
The results of the evaluation of Equation \ref{formation} for the remaining $v_\mathrm{O}$ ground states (gs) of our previous section are depicted in Figure \ref{fig:1V_FE} panel (b) for $\delta_{01}$ and panel (c) for $\delta_{02}$. While most of the gs are semiconductors, there are few configurations which posses a slightly lower energy semi-metallic solution among the several possible initializations we mentioned above. In those cases we have shown both solutions for sake of illustration although the lowest energy solution for what turned out to be the most representative Fe-Co crystal arrangement is a semiconductor state in either $\delta_{01}$ or $\delta_{02}$ cases. Moreover, TM cations such as the magnetic Fe and Co are able to accommodate a wider range of spin states than their diamagnetic counterparts, therefore, any respond through the valence states to stoichiometry conditions could also lead to non-stoichiometry and hopping conductivity \cite{hoppingconductivity}. Nevertheless, mobile charges would screen out the electric polarization, thence, not being suitable for our investigation of the ferroelectric response within the Berry scope. Hence, we sticked with the aforementioned semiconductor states, which were also compared to the gs from hybrid DFT calculations from that viewpoint\cite{STFC_Ox_hybrid}.

Let us here to interpret what Figure \ref{fig:1V_FE} suggested from $E_f$ results. A preference for the $v_\mathrm{O}$ to be coordinated by [100]- and [111]-oriented Fe-Co pairs is observed, as Ti-Ti coordinated vacancies tend to have significantly higher $E_f$. This is congruent with our preference for at least one O$_{5}$ uncompleted octahedra hosting Co and Fe ions in order to reduce $\delta_{02}$ large number of configurations. However, the total energy differences between [100]-configurations and the other two orientations for $\delta_{01}$ and $\delta_{02}$ deficiencies is at least $\sim{0.6}$ eV and $\sim{0.7}$ eV respectively, when comparing [100] and [110], with even larger differences for [111]. Moreover, the energy differences between Fe-Co-arranged pairs favor [100]-configurations by $\sim{0.1}$ vs. [110] and $\sim{0.9}$ eV vs. [111] as Figure \ref{fig:1V_FE}(a) shows. If we intuitively model an experimental process in which the oxygen deposition pressure is progressively tuned e.g., in a PLD synthesis, the resulting O-deficient perovskite would probably have a slightly changing density of vacancies, also influenced by the substrate and surface presences, such that there would be $\delta_{0}$-dominated crystal regions that then give rise to $\delta_{01}$-dominated (and $\delta_{02}$) ones as the pressure changes. Therefore, it is plausible for this experiment to follow the energy trends suggested above such that the gs semiconductor states for the [100] arrangement, which are suggested by hybrid results \cite{STFC_Ox_hybrid}, are the focus of our following analysis. Henceforth, we comprehensively analyzed the magnetic/electric order parameters of the three lowest-energy gs of O-deficient structures with the [100] Fe-Co arrangement and $\delta_{01}$ as pointed out in Figure \ref{fig:1V_FE} (vertical lines). In the case of the same cations arrangement for $\delta_{02}$ we analyzed the three lowest gs plus a fourth gs, which is energetically closest to the mentioned three states as depicted also in Figure \ref{fig:1V_FE}.

The structural and magnetic properties of the representative perovskites selected above are described in Table \ref{tab:1V2V_SmallTable}. The system relaxes to FM ordering in the global ground state with spin-gaps below within $\sim{20}$ meV to AFM orderings in P1 and P4/mmm symmetries for $\delta_{01}$ and $\delta_{02}$ respectively. P1 with the lowest symmetry due to the $v_\mathrm{O}$ perpendicular to the Fe-Co alignment is responsible for the larger u.c. volume respect P4/mmm. The Fe and Co ions are stable as high spin-states for $\delta_{01}$ and $\delta_{02}$, and the magnetic moment/u.c. slightly increases from $[100]_{b}$ to $[100]_{a\text{\textquoteright}}$ as both Fe and Co are sandwiched by $a$ vacancies in ${a\text{\textquoteright}}$. The energy band-gap also increases significantly for any $\delta_{02}$ compared to any $\delta_{01}$ solution. FM and AFM orderings compete among ground solutions depending on the number and location of $v_\mathrm{O}$ with respect to the Fe-Co pair. These qualities are in partial agreement with hybrid calculations \cite{STC,OdefSTF,STFC_Ox_hybrid}. 

\begin{figure*}[t]
\begin{minipage}{1.0\linewidth}
    \includegraphics[width=1.0\columnwidth]{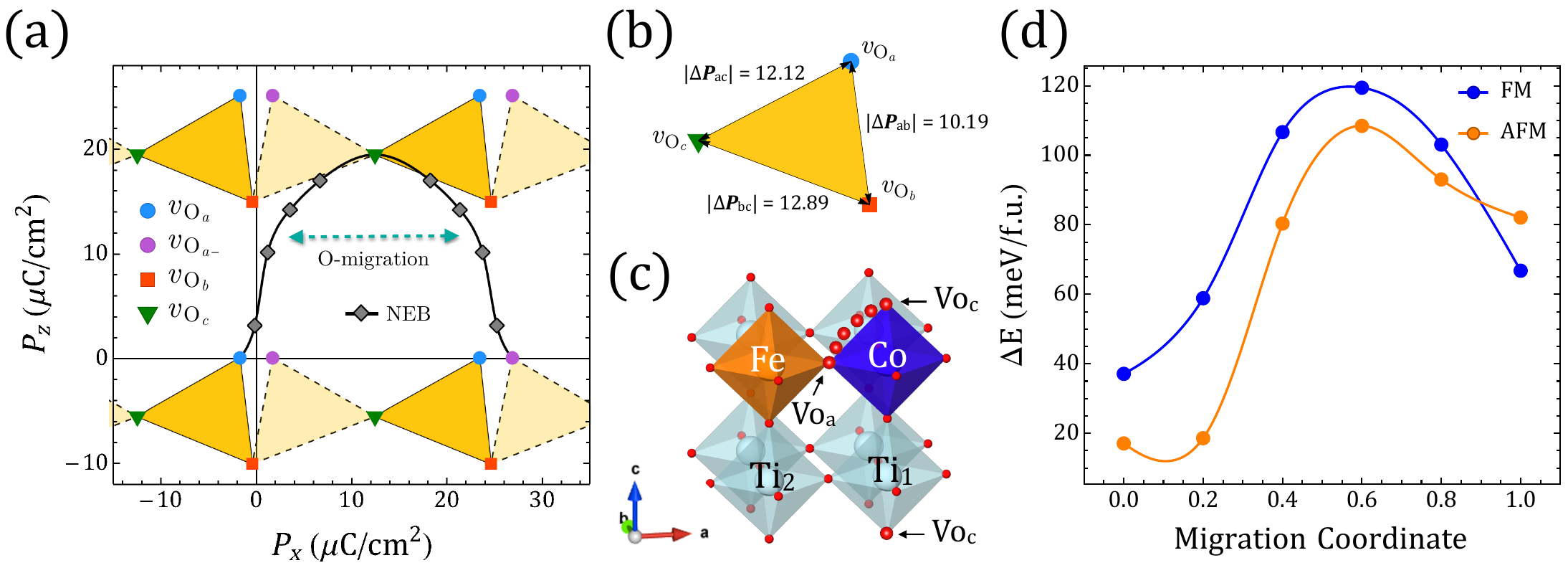}\\
    \vspace{5pt}
    \includegraphics[width=1.0\columnwidth]{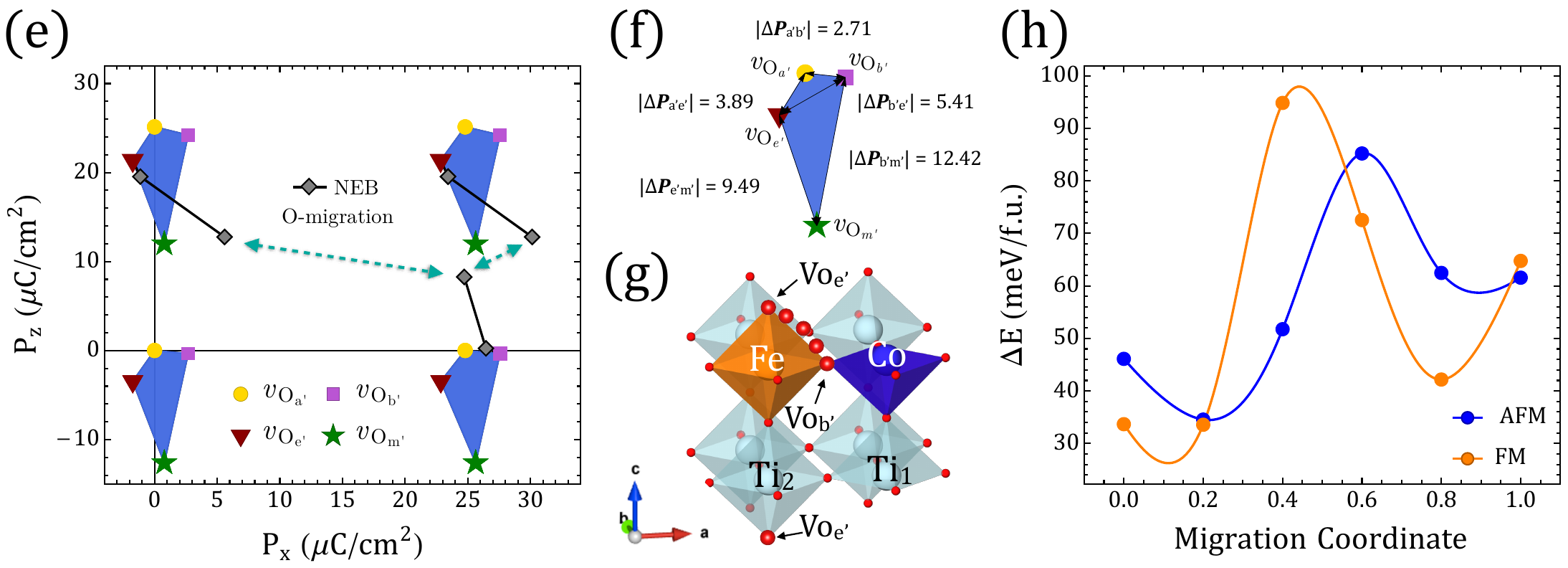}
\end{minipage}
\caption{$\mathbf{P}_{Q}$ differentiated polarization within the Berry's phase approach. (a) $\vert\Delta \mathbf{P}\vert$ between $\delta_{01}$ $gs$ in Table \ref{tab:1V2V_SmallTable}. Symmetrical x-axis $\mathbf{P}$ for $[100]_a$ (dashed) is also shown. (e) Same as (a) but involving $\delta_{02}$ $gs$. (b) The sides of the triangles  represent the minimal $\vert\Delta \mathbf{P}\vert$ for the respective end-states in the inset frame; the $\vert\Delta \mathbf{P}\vert$ values are shown on  each side. (c) STFC supercell with  O-migration between $[100]_a$ $\Leftrightarrow$ $[100]_c$ ($\mathbf{P}$ fingerprint shown in (a)). (f) Same as (c) but between $[100]_{b\text{\textquoteright}}$ $\Leftrightarrow$ $[100]_{e\text{\textquoteright}}$}
\label{fig:1V_Pol}
\end{figure*}

$E_{f}$ in our analysis is not importantly affected by spurious magnetic interactions between vacancies and it’s PBC (periodic-boundary-conditions) images as suggested by HSE calculations in STC and STFC \cite{STFC_Ox_hybrid,STC} i.e., as we shall discuss later, $v_\mathrm{O}$ distributions of interest seem to promote ferroelectric states partially because of the tendency of the $\pi$(d-p) orbitals to localize in lower symmetry structures while the degeneracies are broken principally by breaking of the Hund's rules \cite{JT,CBFerro}, which is suggested by the predominance of high spin states. Also, although negative or negligible $E_f$ have been associated with oxygen mediated phase-separation and non-negligible defect interactions\cite{phases}, our systems stoichiometry and defects configuration are not suitable respect to those phenomena as suggested e.g. by $[111]_{g\text{\textquoteright}}$. Now, we focus in the rest of this paper on studying differences in electrical polarization between the structures illustrated in Figure \ref{fig:1V_Pol} and selected as explained int his section. Global gs as well as higher energy solutions are considered so that magnetization and structural changes can be compared more generally along with the electrical polarization behavior. 
\section{Electric Polarization and O-migration}
The electric polarization order parameter is defined as following:
\begin{equation}\label{eq:pol_total}
	\mathbf{P}= \mathbf{P}_{ion} + \mathbf{P}_{elec}
\end{equation}
with $\mathbf{P}_{ion}$ and $\mathbf{P}_{elec}$ the ionic and electronic contributions respectively, which are such that 
\begin{equation} \label{eq:P_elec_lambda}
	\mathbf{P}_{elec} =  \sum_{n}^{occ} \frac{1}{V_{cell}} \sum_{j=1}^{3} p_{nj} \mathbf{a}_j
\end{equation}
\begin{equation}\label{eq:pol_ion}
	\mathbf{P}_{ion} = \frac{e}{V_{cell}} \sum_i Z_i \mathbf{\tau}_i
\end{equation}
were we have summed over the occupied $\text{n}$ bands, and the ionic contribution is defined by the $Z_i$ charges of the atomic nuclei and their positions $\mathbf{\tau}_i$ in the unit cell.
The ${p_{nj}}$ are the reduced polarizations i.e., the components of the polarization along the primitive lattice vectors, which in terms of the reciprocal lattice vectors have the form
\begin{equation}\label{eq:reduced_pol}
	p_{nj} = \frac{V_{cell}}{2\pi} \mathbf{b}_j \cdot \mathbf{P}_n
\end{equation}
However, as the Bloch functions used to evaluate the Berry connections are not unique\cite{RestaVanderbiltBook,VanderbiltBook,Pol2Dsym,Persistence_vac_20017}, two different sets of periodic functions will derive to the relation 
\begin{equation}
	\widetilde{p}_{nj} = p_{nj}  + e m_j
\end{equation}
where $m_j$ is an integer. Hence, the electric polarization can be expressed as
\begin{equation}\label{eq:pol_quantum}
	\widetilde{\mathbf{P}}_n = \mathbf{P}_n + \frac{e}{V_{cell}} \sum_{j=1}^{3} m_j\mathbf{a}_j = \mathbf{P}_n + \frac{e\mathbf{R}}{V_{cell}} 
\end{equation}
with $\mathbf{R} = m_1\mathbf{a}_1 + m_2\mathbf{a}_2 + m_3\mathbf{a}_3$ a lattice vector. This results implies that the polarization is well defined just up to the module of the $e\mathbf{R}/V_{cell}$ quantity i.e., all the points defined by Equation \ref{eq:pol_quantum} are valid solutions to the Equation \ref{eq:pol_total}. The quantity separating two different polarization values is called the \textit{quatum-of-polarization} $\mathbf{P}_Q$
\begin{equation}
	\mathbf{P}_Q = e\mathbf{R}/V_{cell}
\end{equation}

This quantum is subjected to the indeterminacy introduced by the freedom that we have to chose the atomic base for a periodic lattice of point charges. Moreover, $\mathbf{P}$ is invariant to unitary unit cell translations for which moving the origin of coordinates is not generally a way to resolve the multiple polarization values as it would be the case if such total polarization was a well defined basal property. All this leaves us with a property that can be read from a lattice-of-polarization whose different components along the lattice vectors are expanded by $\mathbf{P}_Q$ such that $\mathbf{P}= \mathbf{P}_{ion} + \mathbf{P}_{elec} + \mathbf{P}_{Q}$.
$\Delta\mathbf{P}$ is only determined within such a quantum uncertainty thence the process of resolving the polarization scheme requires the evaluation at intermediate steps connecting the end-structures of interest; this will result in branches of polarization separated by quantums as displayed in Figure \ref{fig:1V_Pol}\cite{RestaVanderbiltBook,VanderbiltBook,Pol2Dsym,Persistence_vac_20017}. Some of the process that can be used to generate intermediate linking structures could be e.g., allowing ionic displacements, generating defect migrations, applying strain and/or inducing structural changes with pressure or external fields\cite{RestaVanderbiltBook,VanderbiltBook}.

Let us discuss now the evaluation of $\Delta\mathbf{P}$ for the systems in Table \ref{tab:1V2V_SmallTable}. Figure \ref{fig:1V_Pol} (a) and (e) reveal the values of the electrical polarization lattices in STFC for $\delta_{01}$ and $\delta_{02}$, correspondingly. In order to have the change of the polarization, which is the more meaningful quantity from an experimental point of view, we compared such lowest gs using the lattice of polarization given by the different $\mathbf{P}_Q$ solutions, and in that lattice the figures conformed by joining the $\mathbf{P}$ points associated to each defected configuration would have sides of magnitude equivalent to $\vert\Delta \mathbf{P}\vert$. Our first observation is very promising, Figure \ref{fig:1V_Pol} (a) for $\delta_{01}$ shows that O-deficient STFC perovskites can have sufficiently non-centrosymmetric polar structures to yield outstanding $\vert\Delta \mathbf{P}\vert$ values. Figure \ref{fig:1V_Pol} (b) enlarges the smallest possible changes in polarization, within ${e}\mathbf{R}/{V}$ uncertainty, that occurs when the system transitions from $[100]_a$ to $[100]_b$ i.e., 10.19 $\mu$C/cm$^2$; and comparing $[100]_c$-$[100]_b$ or $[100]_a$-$[100]_c$ we obtain 12.89 $\mu$C/cm$^2$ and 12.12 $\mu$C/cm$^2$, respectively. These last values are similar to or higher than that of other ferroelectrics \cite{MultiferroicsSpaldin2019,multiferroico1,otroferrovalue}. Also, these polarization changes in Figure \ref{fig:1V_Pol} (b) are similar when compared PBE+U o HSE calculations. \\

We need a footprint to resolve $\vert\Delta \mathbf{P}\vert$ now, i.e., to find the values of the polarization in between those end-points defining two different structures in Table \ref{tab:1V2V_SmallTable}; those values connect two vertices that might be separated by different $\mathbf{P}_Q$, in which case $\vert\Delta \mathbf{P}\vert$ values would be larger than the ones mentioned above, and would capture features associated to electronic structure changes that differentiate those ground states. Structural deformation due to oxygen diffusion  could lower the symmetry and lead to large lattice-parameter changes as in strained/pressure-induced ferroic transitions, as well as charge transfer and redistribution could be mediated by migrating oxygen \cite{OdefSTF,STC2018,STO111dft}. Hence, we analyze the effects of adiabatic oxygen migration by following NEB-relaxed paths \cite{NEB_polaron,NEB_LSFO} between selected $v_\mathrm{O}$ sites. Figure \ref{fig:1V_Pol} displays two representative migrations while Figures S6-S9\cite{Supplemental} contains complementary ones.\\
\begin{figure}[t]
\includegraphics[width=1.0\linewidth]{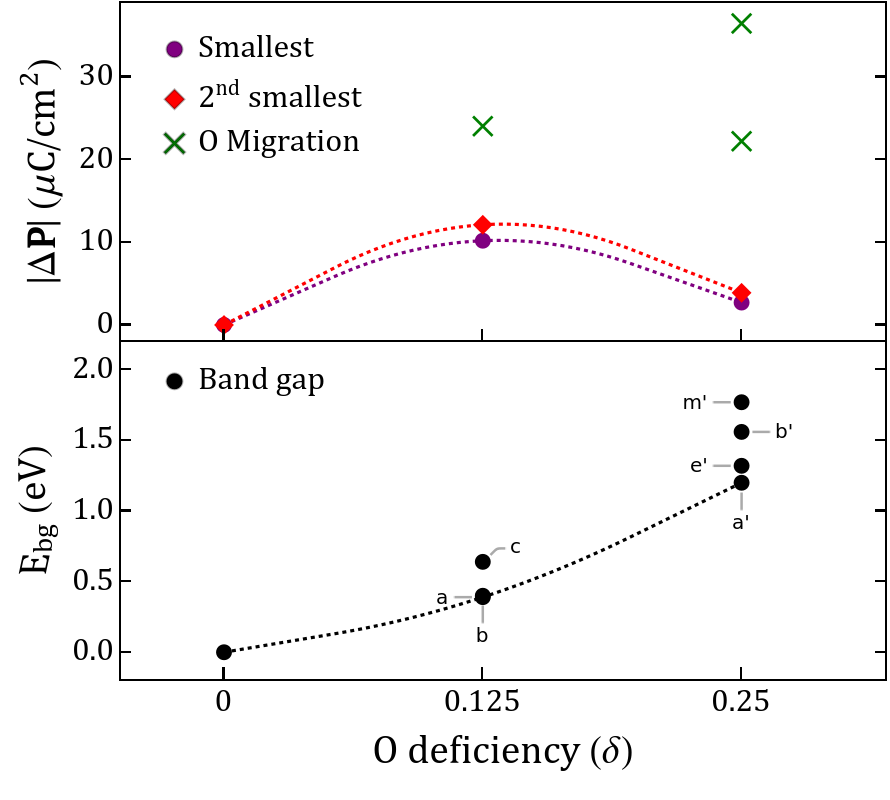}
\caption{Lowest two $\vert\Delta P\vert$ at each $\delta$ (dashed lines), out of all the studied configurations. $\vert\Delta P\vert$ obtained from migrations shown in Figure \ref{fig:1V_Pol} (a) and (b) are marked with crosses. } 
\label{fig:Prop_vs_delta1}
\end{figure}
Symmetry-equivalent $v_\mathrm{O}$ conform symmetric barriers while migrations connecting energetically different are represented by asymmetric barriers as it can be seen in Figure \ref{fig:1V_Pol} and Figures S6-S9\cite{Supplemental}, which resemble e.g., oxygen diffusion along different O$_{5}$-defined NaBiTiO$_{3}$ planes \cite{He:2015dp}. In our case, for instance, symmetric $[100]_{c}$ $v_\mathrm{O}$ have the $z-y$ plane defining $\Delta\mathbf{P}$ as ${P}_{x}$ does not change in that rotation, but the local magnetic moments do not change either because neither Fe or Co coordinations are being effectively modified (Figures S6\cite{Supplemental}). The migrations barriers for symmetric cases are in general comparable or even larger than non-symmetric ones, and in the process of tuning the ferroelectric and/or magnetic order parameters energetically different $v_\mathrm{O}$ are most likely to be involved in a oxygen-driven structural accommodation, which leaves us with the migrations $[100]_{a}$-to/from $[100]_{c}$ or $[100]_{b}$ if we restrict for sake of comparison to $\Delta{E}$ approximated to the separation between the lowest and the second lowest gs. We then displayed in Figure \ref{fig:1V_Pol} (a) the continuous polarization footprint of a system in which an oxygen moves from the Fe-O-Co position to the $v_\mathrm{O}$ in $[100]_c$ and back to the $v_\mathrm{O}$ in $[100]_a$  along the path illustrated in Figure \ref{fig:1V_Pol} (c). Also in Figure S7\cite{Supplemental} the results for migration $a$ to/from $b$ are shown.

In the migration $[100]_{a}$-to $[100]_{c}$ the semiconducting character of the system is preserved throughout the intermediate migration states independently of the spin polarization, and it has a $y$-reflection symmetry and thence provides $P_y=0$ centered solutions for the polarization. The FM path is slightly more expensive than the AFM one though it is cheaper in the vicinity of the $[100]_{c}$ end point Figure \ref{fig:1V_Pol} (d), which means it would require around a spin-gap energy difference to maintain the whole path as a semiconductor one. In turn, the magnetization/u.c. would decrease because of the magnetic order rather than due to the local magnetic moments decrements observed in the FM path (Figure S4\cite{Supplemental}), which is also related to the steepest detriment of the energy band-gap in this last path as TM orbitals occupancy slightly increased; such magnetization is nonetheless of considerable $\sim1\mu_{B}$. The electric polarization on the other hand is now not restricted to the closest values within the $\mathbf{P}$ lattice, such that $\vert\Delta \mathbf{P}\vert$ for $[100]_a$-$[100]_c$ vacancies switching would reach $\sim{24}\mu$C/cm$^2$.\\
\begin{figure}[t]
\includegraphics[width=1.0\linewidth]{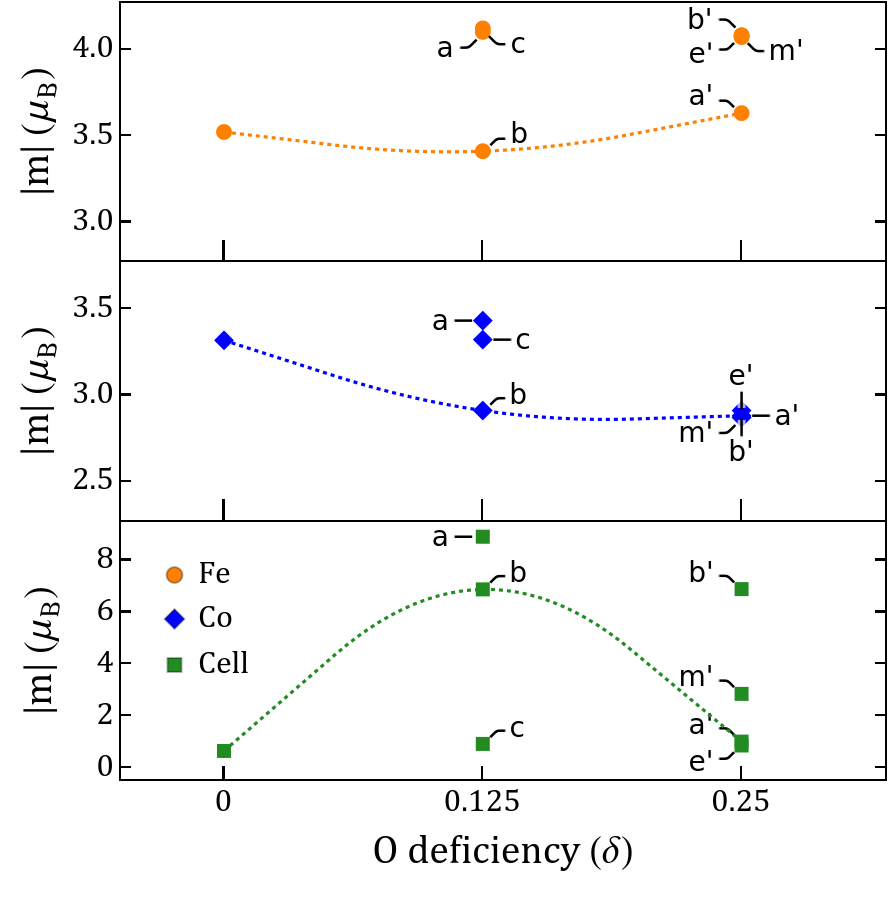}
\caption{Fe/Co and cell magnetic momentvs $\delta$.} 
\label{fig:Prop_vs_delta2}
\end{figure}
Figure \ref{fig:1V_Pol} (e) presents the values of the electric polarization components as well as the $\vert\Delta \mathbf{P}\vert$ values between the $\delta_{02}$ structures in the same way that Figure \ref{fig:1V_Pol} (a) does for $\delta_{01}$ but using instead the four $\delta_{02}$ structures in Table \ref{tab:1V2V_SmallTable}. In that figure we can also see the electric polarization fingerprint for the adiabatic O-migration of the Fe-O-Ti oxygen perpendicular to the Fe-Co line in $[100]_{b\text{\textquoteright}}$ to the Fe-$v_\mathrm{O}$-Co in $[100]_{b\text{\textquoteright}}$, leading to $[100]_{e\text{\textquoteright}}$, as Figure \ref{fig:1V_Pol} (g) illustrates. As we compare figures (b) and (f) is clear that lowest energy gs for $\delta_{02}$ provide us with smaller $\vert\Delta \mathbf{P}\vert$ than those obtained with $\delta_{01}$. We have to take into account an energetically expensive fourth gs $[100]_{m\text{\textquoteright}}$ so we can reach similar polarization values with two vacancies e.g., differences between the first two gs configurations for $\delta_{01}$ and $\delta_{02}$ are 10.19$\mu$C/cm$^2$ and 2.71$\mu$C/cm$^2$, respectively; this is a persistent trend amongst O-deficient perovskites in Figure \ref{fig:1V_FE}. Let us call this the observation (1): connected polar end-structures for $\delta_{01}$ yield larger $\vert\Delta \mathbf{P}\vert$ than those obtained for $\delta_{02}$. 
A second observation (2) is that crystals with such increased $\delta$ yield $\vert\Delta \mathbf{P}\vert$ differences that although are more sensitive to the $v_\mathrm{O}$ symmetry, they are consistently larger when among the three energetically lowest $v_\mathrm{O}$ the two compared $\delta_{02}$ configurations share a $c$ vacancy. Otherwise, sharing no vacancies or sharing $a$ vacancies the changes of polarization are decreasingly lower, respectively. The inclusion of $g$ vacancies will always increases the changes of the polarization but on the contrary to what happen to the lowest three gs, it will favor the duplicity of the $a$ defects. The energy-barrier for $\delta_{02}$ between $[100]_{b\text{\textquoteright}}$ and $[100]_{e\text{\textquoteright}}$ also shows an energy difference between FM and AFM polarizations, which favors the AFM $e\text{\textquoteright}$ polarization throughout the migration in terms of the energy-high but the spin-gap is again relevant in the vicinity of the Fe-O-Ti oxygen position. A third general observation (3) that we can extract from Figure \ref{fig:1V_Pol} and Figures S6-S7\cite{Supplemental} is that for $\delta_{01}$ there seem to be always possible to find an oxygen migration linking two distinct configurations in Table \ref{tab:1V2V_SmallTable} (or their symmetrical representations), which could generate a finite $\Delta \mathbf{P}$ through a continuous polarization-footprint connecting such end-states. For $\delta_{02}$ however, such a polarization-footprint presents a non-monotonic path due to polarization jumps of magnitude comparable to the changes of the polarization resulting from comparing the gs configurations for $\delta_{01}$ and $\delta_{02}$ within the first $\mathbf{P}_{Q}$. While for $\delta_{01}$ uncompleted paths could be found e.g., when some of the intermediate systems during the specific migration turned out to be conductors, in the $\delta_{02}$ case not just all the end-states are semiconductors but they remained as such when disturbed with the vacancy relocation. Therefore, the polarization can always be defined but not always be totally resolved, nevertheless, this means that a change in the branch $\mathbf{P}_{Q}$ separating the available end-solutions of the electric-polarization should always consider at least the closest one as we can see for instance in Figure \ref{fig:1V_Pol}(e), where the migration path is suggesting two different $\vert\Delta \mathbf{P}\vert$ different than the one in Figure \ref{fig:1V_Pol}(f) i.e., $\vert\Delta \mathbf{P}\vert\sim{36} \mu$C/cm$^2$ and $\vert\Delta\mathbf{P}\vert\sim{21} \mu$C/cm$^2$ as can be seem in Figure \ref{fig:Prop_vs_delta1} at $\delta=0.25$. In the following section, we will discuss these jumps of polarization along with several other features of the magnetic/electric response of our oxygen deficient perovskites. 

As it stands now, we have shown so far that they are indeed electrically polarized as well as magnetic, they do display an important change of electric polarization given that we can tune the oxygen vacancy content and/or for a given oxygen deficiency $\delta$ be able to transport oxygen while maintaining the crystal among stable solid-solutions that can be connected with variate energy-barriers depending on the end-states magnetic polarization, furthermore, we have shown that FM or AFM ground states do have a different saturation magnetization but still enough to provide the systems with two field-tunable order parameters. The relation between the magnetic and ferroelectric behavior according to the TM cations (Co/Fe) role, migrations features and the structural and electronic trends is yet to be discussed in the next section along with the picture we should obtain in an intuitive experiment, as the mentioned sections above, in which the oxygen pressure during deposition is increased/decreased in order to look for a specific multiferroic response of the material. 
\begin{figure}[t]
\includegraphics[width=1.0\linewidth]{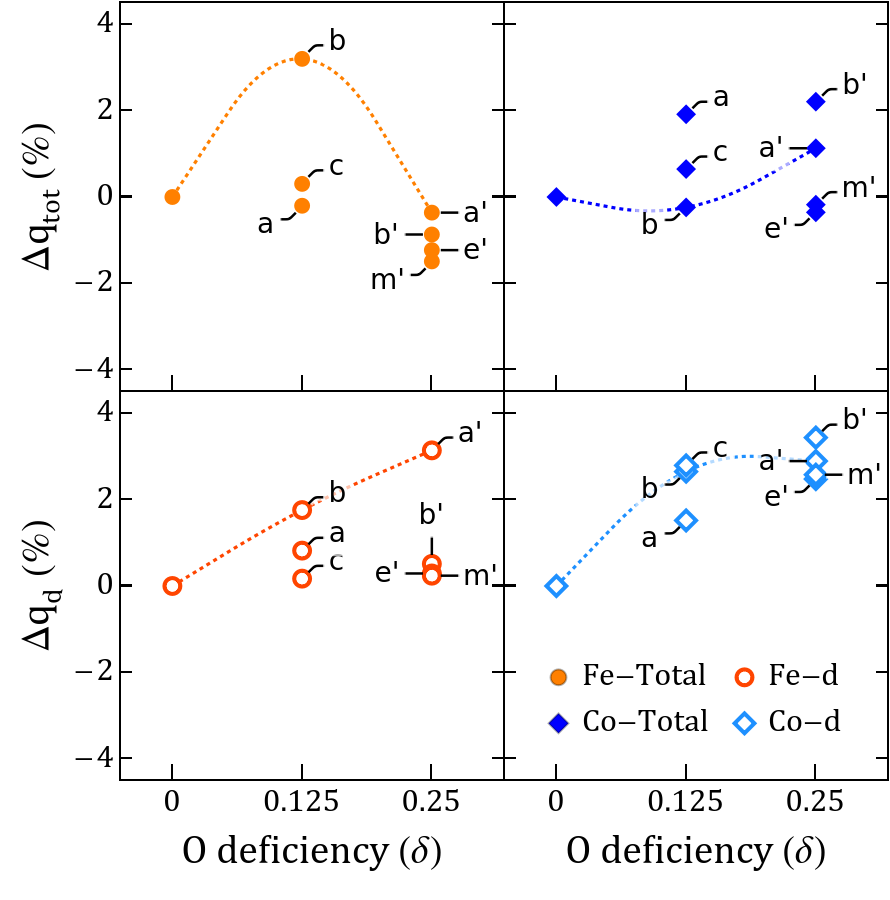}
\caption{Total and d-orbital Fe/Co charge variation vs $\delta$ with respect to $\delta_{0}$ perovskites.} 
\label{fig:Prop_vs_delta3}
\end{figure}
\section{Discussion}
We have discussed these polarization changes in terms of oxygen diffusion. This may be controlled electrochemically, e.g. by ionic liquid gating or oxygen pumping, or  it may be driven by temperature fluctuations or strain effects. This provides the possibility of direct control over the polarizability of such materials post-growth as well as via the oxygen pressure during growth. In Figure \ref{fig:Prop_vs_delta1} we show that while the centrosymmetric stoichiometric solutions do not allow the emergence of a switchable polarization, increasing the deficiency $\delta$ of the perovskite would give us decent polarization changes $\vert\Delta \mathbf{P}\vert$ that are larger for $\delta_{01}$ and decrease for $\delta_{02}$. However, in both cases a migration mechanism can improve upon such a value providing us with at least twice the initial response for one $v_\mathrm{O}$ and up to $\sim{4-6}$ times for two $v_\mathrm{O}$, which is an outstanding possibility. The largest $\vert\Delta \mathbf{P}\vert$ attainable, according to our calculations, likely requires vacancy migration. This mechanism has been widely studied in solid perovskite fuel cells, perovskite-based  capacitors, solid electrolytes and all-oxide electronics \cite{Chroneos:im,deSouza:2015eb}, however, its use in multiferroics as a mechanism to assist ferroic-order is unexplored. \\

\begin{figure}[t]
\includegraphics[width=1.3\columnwidth]{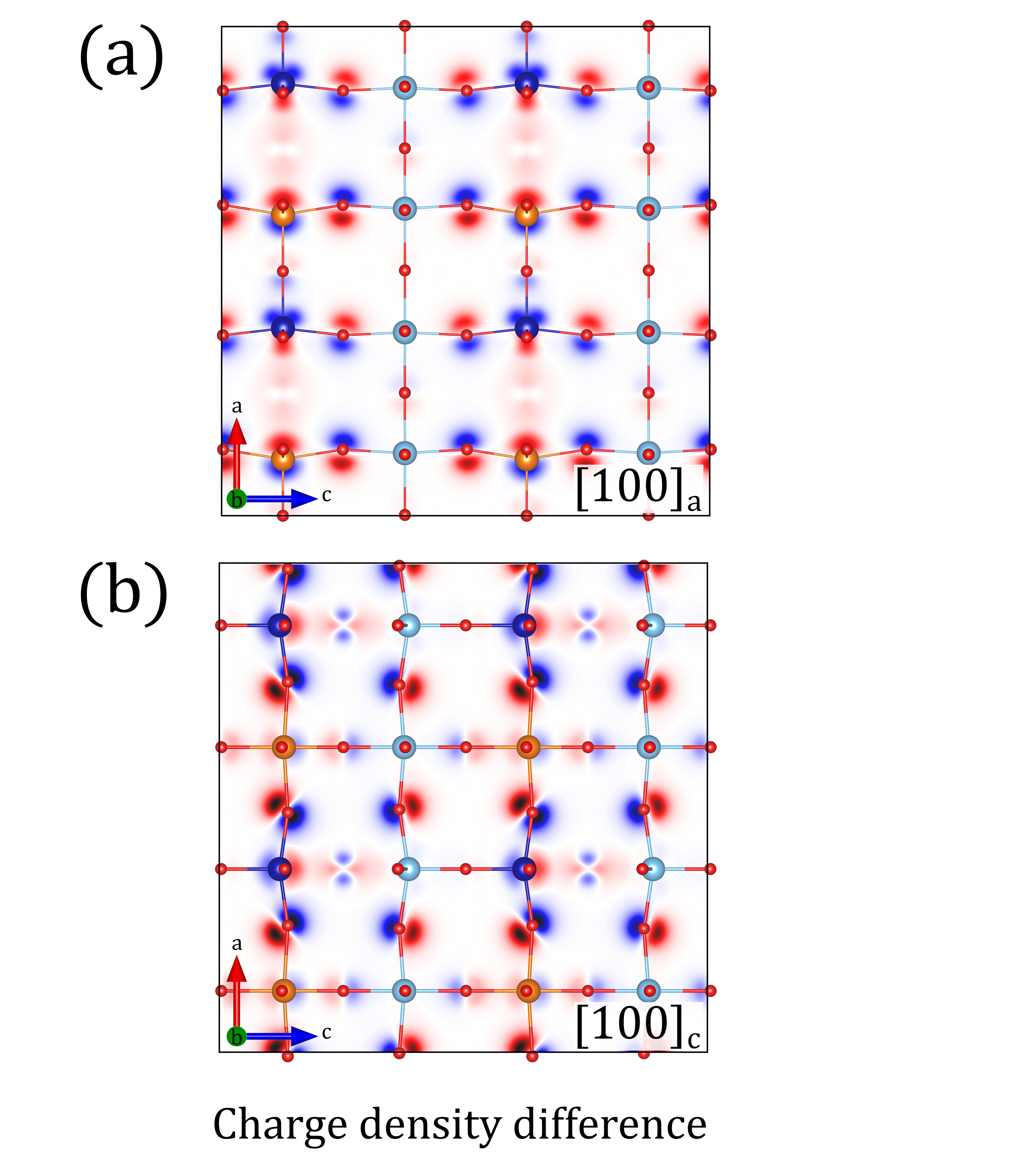}
\caption{Projected charge density difference  $\Delta \rho =\rho(\delta_{01}) - \rho(\delta_{0}) + \rho(O_\text{Removed})$, with $\rho$ the respective charge densities, for $[100]_{a}$ (a) and  $[100]_{c}$ (b) configurations. Charge accumulation/depletion (positive/negative $\Delta \rho$) is represented using blue/red colors}
\label{fig:Chgdiff}
\end{figure}

Figure \ref{fig:Prop_vs_delta2} suggests the magnetization follows a similar behavior as the changes of the electric polarization when $\delta$ increases if we observe it from an absolute gs point of view, nonetheless, if a migration is performed for instance for $\delta_{01}$ and between $a$ and $c$, the magnetization would decrease considerably while the electric polarization switches as described above; if on the other hand $a$ and $b$ are used for an oxygen to migrate the magnetization will slightly decrease while the change of the polarization cannot be determined as the system becomes metallic at intermediate points of the migration. For the migration in Figure \ref{fig:1V_Pol} (e) and (f) at $\delta=0.25$, we will have an inverse change in $\vert\Delta \mathbf{P}\vert$, increasing when relaxing from $e\text{\textquoteright}$ to $b\text{\textquoteright}$, while the magnetization/u.c. decreases. This behavior of the magnetization qualitatively replicates the magnetization trends of experimental results for STF\cite{OdefSTF}. The maximization of both polarization and magnetic moment in some of the cases is at least curious as we primarily have in mind the so-called 3d-orbital paradigm i.e., partly filled 3d-orbitals promote magnetism but disfavor FE \cite{MultiferroicsSpaldin2019,multiferroico1}. 
\begin{figure}[t]
\includegraphics[width=1.0\columnwidth]{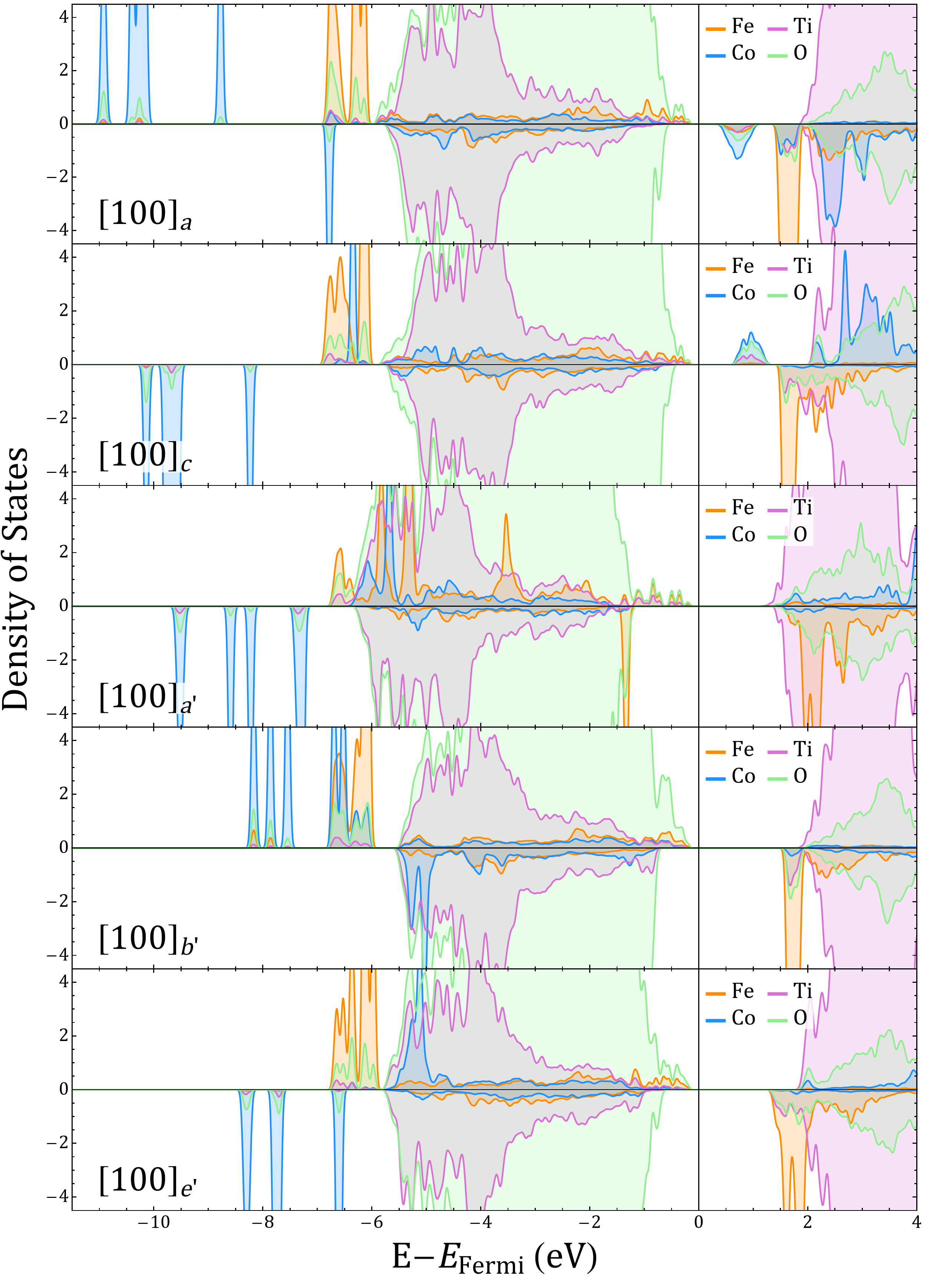}
\caption{Fe, Co, Ti and O projected density of states (DOS) for $[100]_{a,c}$ and $[100]_{a',b',e'}$ corresponding to $\delta_{01}$ and $\delta_{02}$, respectively.}
\label{fig:DOS}
\end{figure}
Let us now to analyze the TM Co and Fe roles through the vacancy arrangements that provide us with the aforementioned behavior. From the point of view of the oxygen mobility comparing vacancies $a$ and $c$ for $\delta_{01}$ is revealing as Figure \ref{fig:Prop_vs_delta2} tells us that the largest change of the magnetic order parameter is not due to a change in the electronic occupancy of both local moments as the vacancy moves between Fe-O$_{6}$/Fe-O$_{5}$ and Co-O$_{5}$, but to the magnetic switching between the two energetically close ground-states. The Co however, has a very small change of its magnetic moment respect to $\delta_{0}$ solutions while Fe has increased more appreciable for $\delta_{01}$, being then responsible for any effective saturation magnetization of the $c$ solution. Figure \ref{fig:DOS} displays the differences between the electronic occupancy of Co/Fe and the cations polarization, this last which is driven by the rotation of the $v_\mathrm{O}$ in the Co O$_{5}$ respect to the Fe-Co line as Figure \ref{fig:Chgdiff} also illustrates. In Figure \ref{fig:Prop_vs_delta3} we can see that such vacancies interchange will modify more importantly the total and d-orbital charges of the Co even though it remains in an uncompleted oxygen octahedra, which is related also to the larger electronegativity of Co respect Fe. Also, the smaller changes for Fe still give rise to a little bit larger increase of the local magnetic moment, which is a sing of an intrinsic charge reorganization among the hybridized orbitals of the Fe-O-Co bonding. \\

Figure \ref{fig:Chgdiff} shows us some of the repercussions of the discussed so far by displaying what happens with $a$ and $c$ for $\delta_{01}$ when looking at the density charge difference with respect to the stoichiometric system. These representations, which are calculated for defected structures with the same symmetry as Table \ref{tab:1V2V_SmallTable} displays, provide us with a simple but powerful idea about the origin of the electric polarization in the perovskites presented here. The resulting structures are non-centrosymmetric and have off-centered Co and Fe cations as well as Ti ions. The uncompleted Fe/Co-O$_{5}$ octahedra present local bending with respect to the plane perpendicular to the Fe-$v_\mathrm{O}$-Co and Ti-$v_\mathrm{O}$-Co lines that is reflected in tetrahedral lattice distortions. The symmetry of the charge density suggests that the creation of the local oxygen vacancy has promoted a subtle charge redistribution that is more evident in the case of Co cations as they present a larger change of local charge, while Fe and Co coordinations have also distinct characteristics beyond the ion radii that are triggered also by the distinct 3d-2p hybridizations of the TM at the Fe-O$_{5}$ and Co-O$_{5}$ centers as it is also suggested by the projected density of states in Figure \ref{fig:DOS} and S10\cite{Supplemental}. One of the signatures of such subtle but distinct Fe and Co behavior can be seen e.g. in Figure \ref{fig:DOS}. For $\delta_{01}$ Co hybridizations define the acceptor-like states delimiting the band-gap while half-filled Fe ones conform the Fermi limit of the valence band. For $\delta_{02}$ nevertheless, the additional electrons partially occupy such acceptor-states, increasing the gap and proving the propensity of Co to fill 3d orbitals under variations of charge, while Fe orbitals tends to reach high-spin states so that the closest empty state has always opposed polarization to the majority populated ones. These features are useful ingredients thinking on the generation of a material whose ferroic degrees of freedom do not exclude each other.\\

With the introduction of a second vacancy, which for sake of comparison is one of the three $a$, $b$ or $c$, we know that the minimal electric polarization changes tend to decrease and $b\text{\textquoteright}$ could maintain a FM magnetization while $e\text{\textquoteright}$ presents a finite AFM one while the Co decreases its local moment and Fe fairly changes. Figure \ref{fig:Prop_vs_delta3} shows that the additional $c$-$v_\mathrm{O}$ mimics what happens with $\delta=0.125$ and Fe decreases even more its charge while Co does the opposite for $b\text{\textquoteright}$, which is congruent with the fact that we have Co-O$_{4}$ and Fe-O$_{5}$ octahedra ($b$ and $c$ are the second and third gs for $\delta_{01}$ as well as $b\text{\textquoteright}$ and $e\text{\textquoteright}$ for $\delta_{02}$, respectively, these last differing in the vacancy switching $a-b$). The Co d-charges changes once again are favored but unexpectedly the Fe d-charges changes are negligible, which comparing $c$ to $e\text{\textquoteright}$ seems rare as Fe-O$_{6}$ passes to a Fe-O$_{5}$ without a Fe-$v_\mathrm{O}$-Co, that for $\delta_{01}$ fosters such a change in favor of Co. Comparing with Figure \ref{fig:Prop_vs_delta2} and the projected density of states in Figures S10\cite{Supplemental} is clear then that the Fe/Co t$_{2g}$e$_{g}$ electron population is not longer defining the saturation magnetization alone but the previously magnetically neutral Ti are able now to play a role by contributing with a magnetization equivalent to one electron, as seen in Figure S5\cite{Supplemental}, where the magnetization originates on the O$_{5}$-coordinated Ti$_{1}$ and Ti$_{2}$ for Co-$v_\mathrm{O}$-Ti in $[100]_{b\text{\textquoteright}}$ and Fe-$v_\mathrm{O}$-Ti in $[100]_{e\text{\textquoteright}}$ respectively.\\

Ti$_{2}$ for instance, as we can see in Figure \ref{fig:1V_Pol} and Figure S5\cite{Supplemental} for migrations between $b\text{\textquoteright}$ and $e\text{\textquoteright}$, passes from zero to a finite $\sim{1\mu_{B}}$ half way the migration path; Ti$_{1}$ behaves the oppose of that. This happens when Co-coordination passes from O$_{4}$ to O$_{5}$ while Fe-coordination remains O$_{5}$. In this last configuration a Co-O-Ti$_{1}$ 3d-2p hybridization turns Ti$_{1}$ off with that additional Co charge as Figures \ref{fig:Prop_vs_delta3} and Figure \ref{fig:DOS} show, while an oxygen migrates to form $a$, the Ti octahedra can intuitively be thought as if there is a charge transfer process in which Ti$_{2}$ has now a similar hybridization to the mentioned before but in this case for Fe-O-Ti$_{2}$, with the Ti now in an uncompleted Ti-O$_{5}$. This magnetization process of Ti ions through hybridised orbitals that represent a process in which an electron is given and received thanks to the different electronegativities and electronic valences as well as to the local defect topology respect to the neighboring TM cations is a little different to what happens in vacancy induced magnetism in  SrTiO$_{3-\delta}$ \cite{STO_Review} as in that case non-filled 3d-Ti orbitals can locally define the magnetization. The Ti$_{1,2}$ magnetic activation in $b\text{\textquoteright}$ and $e\text{\textquoteright}$ is then due to a  superexchange-like mechanism between those ions that is dominated by the Co and Fe electronic environment response to the $v_\mathrm{O}$.

\begin{figure}[t]
\includegraphics[width=0.7\columnwidth]{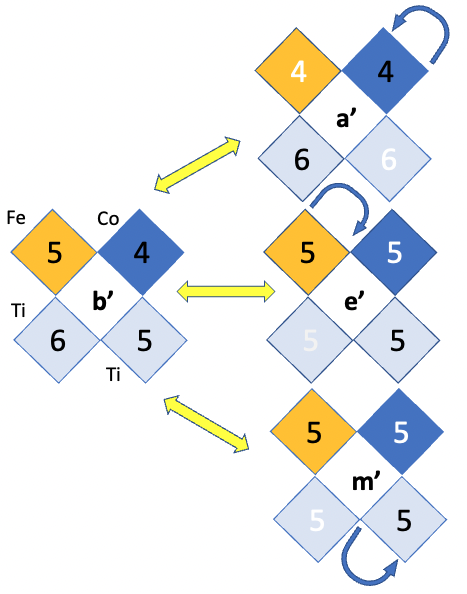}
\caption{Coordinations of the Ti, Co and Ti$_{1,2}$ ions as the the migrations connecting $b\text{\textquoteright}$ with $a\text{\textquoteright}$, $e\text{\textquoteright}$ and $m\text{\textquoteright}$ occur. Arrows point out the vertex between which the oxygen move starting in this case e.g, from $b\text{\textquoteright}$. Black numbers represents O coordinations that do not change from $b\text{\textquoteright}$ to any other configuration while whites ones have changed by one.}
\label{fig:Coor}
\end{figure}

To see how this mechanism works from an intuitive charge-transference viewpoint associated to the relocation of the vacancies, which is equivalent to the change in the TM octahedra coordinations, we can follow a pictoric description in Figure \ref{fig:Coor} as its caption describes. In the case of the migration in Figure \ref{fig:1V_Pol} the Fe and Ti$_{1}$ remain in 5 coordination but Co and Ti$_{2}$ change from 4 to 5 and 6 to 5, respectively. So, a Ti coordination changed by one means an effective electron charge loss or gain, which mean a $\mp1{\mu_{B}}$ titanium magnetization change. On the other hand, such change in the local magnetic TM will change the TM valence spin state, as already discussed here and other cited references, but also an effective $\pm1{\mu_{B}}$ variation in the Ti will be the result of the aforementioned hybridizations, as the covalent character of the Co/Fe-O-Ti$_{1,2}$ persists. The initial polarization of the Ti ions respect to the magnetic cations follows the magnetic description here and in references \cite{OdefSTF,STC,STFC_Ox_hybrid}, which puts the Ti$_{1,2}$ ions with a different polarization respect to Co for Co parallel to Fe and the opposed of that for AFM coupling. The magnetic solutions for the crystal images in between the end-states will simply switch as can be seen appreciated in Figure S5\cite{Supplemental} for all the $\delta_{02}$ migrations.
This behavior just described is common among all the $\delta_{02}$ systems we handled in this investigation.\\
Figures S6-S9\cite{Supplemental} show that migrations barriers in STFC range between $\sim 40-130$ meV/f.u., the lowest values obtained for migrations involving "excited" vacancies coordinating Ti ions. Comparatively, some of the previously calculated ones for STO are around $\sim{50}$ meV/f.u. \cite{Anonymous:2016eh}. Our energy barriers should be taken as witnesses of the possible costs of an oxygen mobility process, however, as they are measured respect the global gs that we have in Table \ref{tab:1V2V_SmallTable} and connect end-states that can have slightly different symmetry, features as the small well-like steps separating the large barrier from the stable gs are attributed mostly to the relaxation of the O$_{5,4}$ uncompleted octahedra whose bending symmetry respect the Co/Fe-$v_\mathrm{O}$ line is broken during the migration. O$_{5}$ undergoes an initial rigid rotation before bending the planes perpendicular to the (Fe/Co)-$v_\mathrm{O}$. This asymmetric energy landscape resembles the trapping effect in Gd-doped CeO$_{2}$ \cite{Vives:2015dd} which however, can be tuned through oxygen vacancy migrations. Pronounced features as such that could even reach negative energy in few cases can also be representing the instability of the intermediate states when a conductor-semiconductor transition is possible as Figures S7\cite{Supplemental} shows. Although the final path of the oxygen could be leading the ions into otherwise inter-octahedra space, as displayed by the z-(x,y) solutions for the NEB coordinates in Figures S6\cite{Supplemental}, this is not really an interstitial phenomena as observed in electrolytes under strain \cite{DeSouza:2012jl} or in promotion of TM shifting/diffusion through $v_\mathrm{O}$ migration in bulk LiNiO$_{2}$ \cite{Kong:2019gm}. In this last case and from our scope, there is always a spin polarization solution that keeps the systems as a semiconductor. In general, oxygen migration between O$_{4,5,6}$ sites depends on the pathway and the coordinating A(B) ions for the end-$v_\mathrm{O}$ as we have three different B-ions stabilized in different spins states, so we expect our systems to mimic a variety of features for the migration barriers as  described for O$_{2}$ diffusion in NaBiTiO$_{3}$ \cite{He:2015dp} without loss of generality respect to the ferroic order parameters changes, which still depend upon the full relaxed gs.\\

Also, lattice distortions in Figure \ref{fig:Chgdiff} and Table \ref{tab:1V2V_SmallTable} suggest that $\vert\Delta \mathbf{P}\vert$ for $\delta_{02}$-configurations could be enhanced with lattice strain, which is useful if we were to use strain mechanisms to tune the migration barriers while conserving specific symmetries for the defected end-structures \cite{DeSouza:2012jl}. Vacancies structural effect by itself can produce or suppress polar structures, e.g. $[100]_{a}$ has a P4mm structure without inversion symmetry along the Fe-Co direction while $[100]_{c}$ has a Pmm2 orthorhombic structure, they both represent a finite change of the electric polarization as seen in Figure \ref{fig:1V_Pol}. In the gs case for $\delta_{02}$, the inclusion of a second $v_\mathrm{O}$ in O-Fe-$v_\mathrm{O}$-Co reestablishes aforementioned inversion symmetry and as a consequence the polarization decreases. Nonetheless, any $\vert\Delta \mathbf{P}\vert$ still requires minimum end-structures optical properties, in this case a finite band-gap. Although $\delta_{02}$ systems are semiconductors, the migrations paths in Figure \ref{fig:1V_Pol} and Figures S8-S9\cite{Supplemental} seem to be unable to completely resolve $\vert\Delta \mathbf{P}\vert$ even when such semiconductor behavior remains in the intermediate structures. This is because of the jumps in the polarization fingerprint, which follow the charge transference that separates two end-states as Figure \ref{fig:DOS} and supplementary figures illustrate. A denser grid of intermediate structures would still have some jumps in the path as redistributed charges into ions that are in generally not symmetrically equivalent are the responsible for the largest part of this polarization changes. The magnetization though could be used to narrow the structural region of imminent electronic transference by using for instance a climbing point technique, but this is out of our scope.\\

The ideas we have discussed about modulating the order parameters by tuning the oxygen deficiency as well as promoting oxygen mobility in deficient structures should not be far from being realized as such mechanisms are being recently exploited in variety of systems. Experimental engineering of oxygen-deficient TM-based magnets/ferroelectrics is now within reach. Stable three-state nonvolatile memory devices were realized by combining both ferroelectricity and oxygen vacancy migration in Pt/BTO/STO, were oxygen vacancies modify the switching properties \cite{Lu:eb}. Combined in situ scanning probe and transmission electron microscopy has been used to study the field-induced migration of oxygen vacancies in thin films of PrCaMnO$_{3}$. In this last case the existing oxygen vacancies in the material have been imaged in situ  and are found to migrate under an external electric field \cite{Liao:ki}.
Measurements of thermally stimulated and pyroelectric currents were performed in STO single crystals subjected to an electric field. A dielectric to pyroelectric phase transition in an originally centrosymmetric crystal structure with an inherent dipole moment is found, which is induced by oxygen migration \cite{Hanzig:ch}.
Moreover, changes in oxygen content are achievable e.g., by ionic liquid gating, so one has a path to modulating the properties in real time \cite{Leighton:2019bo}. For instance, it has been recently shown that redox-driven reversible topotactic transformation in epitaxial SrFe$_{0.8}$Co$_{0.2}$O$_{3-\delta}$ thin films can be achieved at low temperature and at atmospheric pressure. This transformation triggers changes in electronic structures as well. These reversible redox reactions and/or associated changes at low temperature and under atmospheric pressure are particularly valuable to develop a cathode for  solid-oxide fuel cells \cite{Lee:2018dr}.\\
\section{Conclusions}
To summarize, we demonstrated that STFC is a magnetic semiconductor capable of sustaining electric polar structures for a range of TM orderings and O-deficiencies. Variations of $\delta$ and O-migration are effective mechanisms to tune electric and magnetic polarization changes and therefore engineer perovskites using O-deficiency and cation arrangement. 

The Fe and Co TM contribute distinctly through their electronegativity, radii and spin valences to the order-parameters response to the oxygen vacancies, which allow us to tailor variety of vacancy densities and distributions such that both magnetic and ferroelectric orderings can be enhanced. 

The preferred ground states of STFC convey layered perovskites with sheets of linear organized O$_{4,5,6}$-coordinated Fe-Co pairs, separate by Ti ions, and sandwiched with layers of O$_{5}$-coordinated Ti, which is a first suggestion of our model for the engineering of the STFC-based multiferroic. The model suggests that $v_\mathrm{O}$ are not uniformly spread all over the crystal but they stabilize at the Fe-Co octahedra following their arraignment, which leaves us once again with a resulting layered STFC in which the magnetic TM are locally deficient in contrast to the Ti layers that are mostly not defected. The Co,Fe-$v_\mathrm{O}$ defects are the source of the crystal symmetry distortions, off-centering of cations and bending of the oxygen octahedra. Theses are hints for the experimental techniques to be applied such that these layered-like STFC could be realized.

Deficiency according to one $v_\mathrm{O}$/u.c. provided us with the largest values for the changes of the magnetic and electric parameters, though there are several stabilized structures that in the worst case scenario would always display a small but finite saturation magnetization while the polarization changes are very similar among those structures. On the other hand, deficiency according to two $v_\mathrm{O}$/u.c. would usually provide us with smaller electric polarization changes as well as smaller saturation magnetizations, but there are several stabilized structures that could increase such magnetization upon the most likely relaxed-structure one while a second source of magnetization appears as the Ti layers could also become magnetic.


Oxygen migrations for $\delta=0.125$ provides us with $\vert\Delta \mathbf{P}\vert$ $>\sim10 \mu$C/cm$^2$ due to a quantum-of-polarization differences between $\delta=0.125$ structures. Increasing the deficiency to $\delta=0.25$ yields $\vert\Delta \mathbf{P}\vert$ whose O-migration resolved polarization for $\delta=0.25$ is $>\sim3 \mu$C/cm$^2$ in the worst case scenario. All those values of electric polarization changes are considerable larger compared to any STO-based multiferroic, however, the electric order parameter should be further tested respect reversibility and hysteresis so a ferroelectricity is more properly declared under a practical scope.
It is important to mention that the ferroic order parameters studied here shall not be included yet into any defined group among ferroelectric materials such as proper or improper ones. The source of the magnetism and the electric polarization in term of the magnetoelectric coupling should be studied within another scope, for now, this path to multiferroism is new as far as the authors knowledge.\\

We have presented physics in the form of an intuitive set of rules for STFC to be multiferroic, but such rules are expected to be a particular case of a broader  set for magnetic oxides, which are based on the interplay between (Co/Fe)-$v_\mathrm{O}$ defects, TM-cation coordination and topological vacancy disorder. The manipulation of both ferroelectricity and oxygen vacancies according to such rules could facilitate applications such as non-volatile memory, whereas  oxygen content-driven phase transitions will provide a path  to tune the physical properties as well as an important strategy for low-temperature mixed TM electronics.\\

J. M. Florez thanks project PI$_{-}$LIR$_{-}$2021$_{-}$100 from DGIIE-USM and E. A. Cort\'es thanks DGIIP Master scholarship and PIIC-031/2017. S. P. Ong acknowledges support from the Materials Project, funded by the U.S. Department of Energy, Office of Science, Office of Basic Energy Sciences, Materials Sciences and Engineering Division under contract no. DE-AC02-05-CH11231: Materials Project program KC23MP. C. A. Ross acknowledges support from NSF award DMR 1419807.


\section*{References}
\begin{thebibliography}{100}
\bibitem{MultiferroicsSpaldin2019} N. A. Spaldin and R. Ramesh, Advances in magnetoelectric multiferroics, Nat. Mater. 18, 203 (2019)

\bibitem{multiferroico1} M. Fiebig, T. Lottermoser, D. Meier, and M. Trassin, The evolution of multiferroics, Nat. Rev. Mater. 1, 16046 (2016)

\bibitem{MnPerovskite_interplay} A. Marthinsen, C. Faber, U. Aschauer, N. A. Spaldin, and S. M. Selbach, Coupling and competition between ferroelectricity, magnetism, strain, and oxygen vacancies in AMnO$_3$ perovskites, MRS Commun. 6, 182 (2016)

\bibitem{OdefSTF} T. Goto, D. H. Kim, X. Sun, M. C. Onbasli, J. M. Florez, S. P. Ong, P. Vargas, K. Ackland, P. Stamenov, N. M. Aimon, M. Inoue, H. L. Tuller, G. F. Dionne, J. M. Coey, and C. A. Ross, Magnetism and Faraday Rotation in Oxygen-Deficient Polycrystalline and Single-Crystal Iron-Substituted Strontium Titanate, Phys. Rev. Applied 7, 024006 (2017)

\bibitem{STC} J. M. Florez, S. P. Ong, M. C. Onbasli, G. F. Dionne, P. Vargas, G. Ceder, and C. A. Ross, 
First-principles insights on the magnetism of cubic SrTi$_{1-x}$Co$_x$O$_{3-\delta}$, Appl. Phys. Lett. 100, 252904 (2012)

\bibitem{STC2018} A. S Tang, M. C. Onbasli, X. Sun, and C. A. Ross, Thickness-Dependent Double-Epitaxial Growth in Strained SrTi$_{0.7}$Co$_{0.3}$O$_{3-\delta}$ Films, ACS Appl. Mater. Interfaces, 10, 7469 (2018)

\bibitem{STO_Review}  Y. Pai, A. Tylan-Tyler, P. Irvin, and J. Levy, Physics of SrTiO$_3$-based heterostructures and nanostructures: a review, Rep. Prog. Phys. 81, 036503 (2018)

\bibitem{STO_comp_AFD-FE} U. Aschauer and N. A. Spaldin, Competition and cooperation between antiferrodistortive and ferroelectric instabilities in the model perovskite SrTiO$_3$, J. Phys.: Condens. Matter 26, 122203 (2014)

\bibitem{STO_Efield} J. Hemberger, P. Lunkenheimer, R. Viana, R. B\"ohmer, and A. Loidl, Electric-field-dependent dielectric constant and nonlinear susceptibility in SrTiO$_3$, Phys. Rev. B 52, 13159 (1995)

\bibitem{STO_RTFE_strain} J. H. Haeni, P. Irvin, W. Chang, R. Uecker, P. Reiche, Y. L. Li, S. Choudhury, W. Tian, M. E. Hawley, B. Craigo, A. K. Tagantsev, X. Q. Pan, S. K. Streiffer, L. Q. Chen, S. W. Kirchoefer, J. Levy, and D. G. Schlom, Room-temperature ferroelectricity in strained SrTiO$_3$, Nature 430, 758 (2004)

\bibitem{STO_RTFE_tetra} Y. S. Kim, D. J. Kim, T. H. Kim, T. W. Noh, J. S. Choi, B. H. Park, and J.-G. Yoon, Observation of room-temperature ferroelectricity in tetragonal strontium titanate thin films on SrTiO$_3$ (001) substrates, Appl. Phys. Lett. 91, 042908 (2007)

\bibitem{STO_strainfree_FE} H. W. Jang, A. Kumar, S. Denev, M. D. Biegalski, P. Maksymovych, C. W. Bark, C. T. Nelson, C. M. Folkman, S. H. Baek, N. Balke, C. M. Brooks, D. A. Tenne, D. G. Schlom, L. Q. Chen, X. Q. Pan, S. V. Kalinin, V. Gopalan, and C. B. Eom, Ferroelectricity in Strain-Free SrTiO$_3$ Thin Films, Phys. Rev. Lett. 104, 197601 (2010)

\bibitem{STO_Ti_antisite} M. Choi, F. Oba, and I. Tanaka, Role of Ti Antisitelike Defects in SrTiO$_3$, Phys. Rev. Lett. 103, 185502 (2009)

\bibitem{STO_SrOO_vac}  Y. S. Kim, J. Kim, S. J. Moon, W. S. Choi, Y. J. Chang, J.-G. Yoon, J. Yu, J.-S. Chung, and T. W. Noh, Localized electronic states induced by defects and possible origin of ferroelectricity in strontium titanate thin films, Appl. Phys. Lett. 94, 202906 (2009)

\bibitem{STO_Pol_defects} K. Klyukin and V. Alexandrov, Effect of intrinsic point defects on ferroelectric polarization behavior of SrTiO$_3$, Phys. Rev. B 95, 035301 (2017)

\bibitem{STO_FE_SrTi_ratio}  F. Yang, Q. Zhang, Z. Yang, J. Gu, Y. Liang, W. Li, W. Wang, K. Jin, L. Gu, and J. Guo, Room-temperature ferroelectricity of SrTiO3 films modulated by cation concentration, Appl. Phys. Lett. 107, 082904 (2015)

\bibitem{STO001} I. Sokolovic, M. Schmid, U. Diebold, and M. Setvin, Incipient ferroelectricity: A route towards bulk-terminated, Phys. Rev. Mater. 3, 034407 (2019)

\bibitem{STO111} I. Hallsteinsen, M. Nord, T. Bolstad, P.-E. Vullum, J. E. Boschker, P. Longo, R. Takahashi, R. Holmestad, M. Lippmaa, and T. Tybell, Effect of Polar (111)-Oriented SrTiO$_3$ on Initial Perovskite Growth, Cryst. Growth Des. 16, 2357 (2016)

\bibitem{ferro1} Xu, T., Shimada, T., Araki, Y., Wang, J. \& Kitamura, T. Multiferroic Domain Walls in Ferroelectric PbTiO3 with Oxygen Deficiency. Nano Letters 16, 454–458 (2015). 

\bibitem{ferro2} Lee, K. et al. Enhanced ferroelectric switching speed of Si-doped HfO2 thin film tailored by oxygen deficiency. Sci Rep-uk 11, 6290 (2021).

\bibitem{ferro3} Cheng, S. et al. Manipulation of Magnetic Properties by Oxygen Vacancies in Multiferroic YMnO3. Adv Funct Mater 26, 3589–3598 (2016). 

\bibitem{ferro4} Agrawal, P. et al. Strain-driven oxygen deficiency in multiferroic SrMnO3 thin films. Phys Rev B 94, 104101 (2016). 

\bibitem{ferro5} Glinchuk, M. D. et al. Ferroelectricity induced by oxygen vacancies in relaxors with perovskite structure. 1–8 (2018) doi:10.1103/physrevb.98.094102. 

\bibitem{ferro6} El-Naser, A. A. et al. Study the influence of oxygen-deficient ($\delta$ = 0.135) in SrFeO$_{3-\delta}$ nanoparticles perovskite on structural, electrical and magnetic properties. Philos Mag 101, 1–19 (2020).
  
\bibitem{STF_multiferroic} Y.-G. Wang, X.-G. Tang, Q.-X. Liu, Y.-P. Jiang, and L.-L. Jiang, Room Temperature Tunable Multiferroic Properties in Sol-Gel-Derived Nanocrystalline Sr(Ti$_{1-x}$Fe$_x$)O$_{3-\delta}$ Thin Films, Nanomaterials 7, 264 (2017)

\bibitem{STF_multiferroic2} X. Wang, Z. Wang, Q. Hu, C. Zhang, D. Wang, and L. Li, Room temperature multiferroic properties of Fe-doped nonstoichiometric SrTiO$_3$ ceramics at both A and B sites, Solid State Commun. 289, 22 (2019)

\bibitem{APS}  J. M. Florez, M. C. Onbasli, D. H. Kim, S. P. Ong, G. Ceder, P. Vargas, and C. A. Ross, Abstract: M32.00014, APS March Meeting 2015, Volume 60, Number 1 (2015)

\bibitem{Sluchinskaya:2019hu} I. A. Sluchinskaya and A. I. Lebedev, Cobalt in Strontium Titanate as a New Off-Center Magnetic Impurity, Phys. Solid State 61, 390 (2019)

\bibitem{STFC_Ox_hybrid} M. A. Opazo, S. P. Ong, P. Vargas, C. A. Ross, and J. M. Florez, Oxygen-vacancy tuning of magnetism in SrTi$_{0.75}$Fe$_{0.125}$Co$_{0.125}$O$_{3-\delta}$ perovskite, Phys. Rev. Mater., 3, 014404 (2019)

\bibitem{STFC_Ox_memb} Y. Liu, S. Baumann, F. Schulze-K\"uppers, D. N. Mueller, O. Guillon, Co and Fe co-doping influence on functional properties of SrTiO$_3$ for use as oxygen transport membranes, J. Eur. Ceram. Soc. 38, 5058 (2018)

\bibitem{stobased_applications}  B. L. Phoon,  C. W. Lai, J. C. Juan, P. L. Show and  W. H. Chen, A review of synthesis and morphology of SrTiO$_3$ for energy and other applications, Int. J. Energy Res., 43, 5151 (2019)

\bibitem{solidcellsstfcbased} A. Mrozi\'nski, S. Molin, J. Karczewski, T. Miruszewski, and P. Jasi\'nski, Electrochemical properties of porous Sr$_{0.86}$Ti$_{0.65}$Fe$_{0.35}$O$_{3}$ oxygen electrodes in solid oxide cells: Impedance study of symmetrical electrodes, Int. J. Hydrog. Energy, 44, 1827 (2019)

\bibitem{STF_molecular} Kim, D. H. et al. Magnetostriction in epitaxial SrTi1-xFexO3-delta perovskite films with x=0.13 and 0.35. Journal of Physics-Condensed Matter 25, (2013)

\bibitem{SCF_voltage} S. Ning, Q. Zhang, C. Occhialini, R. Comin, X. Zhong and C. A. Ross, Voltage Control of Magnetism above Room Temperature in Epitaxial SrCo$_{1-x}$Fe$_{x}$O$_{3-\delta}$, ACS Nano, 14, 8949 (2020)

\bibitem{vasp3} G. Kresse and J. Furthm\"uller, Efficient iterative schemes for \textit{ab initio} total-energy calculations using a plane-wave basis set, Phys. Rev. B 54, 11169 (1996)

\bibitem{PAW} G. Kresse and D. Joubert, From ultrasoft pseudopotentials to the projector augmented-wave method, Phys. Rev. B 59, 1758 (1999)

\bibitem{NEB} H. Jonsson, G. Mills and K. W. Jacobsen, Nudged elastic band method for finding minimum energy paths of transitions. In: Classical and Quantum Dynamics in Condensed Phase Simulations, ed. B. J. Berne, G. Ciccotti and D. F. Coker, World Scientific (1998) 

\bibitem{ldau_vasp} A. Rohrbach, J. Hafner and G. Kresse, Electronic correlation effects in transition-metal sulfides, J. Phys.: Condens. Matter 15, 979 (2003)

\bibitem{ModPol4} R. Resta, Macroscopic polarization in crystalline dielectrics: the geometric phase approach, Rev. Mod. Phys. 66, 899 (1994)

\bibitem{VESTA} K. Momma and F. Izumi, VESTA 3 for Three-Dimensional Visualization of Crystal, Volumetric and Morphology Data, J. Appl. Cryst. 44, 1272 (2011)

\bibitem{crystalMaker}  CrystalMaker Software Ltd, Oxford, England (www.crystalmaker.com)

\bibitem{pymatgen} S. P. Ong, W. D. Richards, A. Jain, G. Hautier, M. Kocher, S. Cholia, D. Gunter, V. L. Chevrier, K. A. Persson, and G. Ceder, Python Materials Genomics (pymatgen): A robust, open-source python library for materials analysis, Comput. Mater. Sci. 68, 314 (2013)

\bibitem{Assat:2018hx} G. Assat and J.-M. Tarascon, Fundamental understanding and practical challenges of anionic redox activity in Li-ion batteries, Nat. Energy 3, 373 (2018) 

\bibitem{Myeong:2018gd} S. Myeong, W. Cho, W. Jin, J. Hwang, M. Yoon, Y. Yoo, G. Nam, H. Jang, J.-G. Han, N.-S. Choi, M. G. Kim and J. Cho, Understanding voltage decay in lithium-excess layered cathode materials through oxygen-centred structural arrangement, Nat. Commun. 9, 3285 (2018) 

\bibitem{vaspbackground} G. Makov and M. C. Payne, Periodic boundary conditions in \textit{ab initio} calculations, Phys. Rev. B 51, 4014 (1995)

\bibitem{roomt1} D. H. Kim, L. Bi, P. Jiang, G. F. Dionne, and C. A. Ross, Magnetoelastic effects in SrTi$_{1-x}$M$_{x}$O$_{3}$ (M = Fe, Co, or Cr) epitaxial thin films, Phys. Rev. B 84, 014416 (2011)

\bibitem{Supplemental} See Supplemental Material at [URL will be inserted by publisher] for the description of $\delta_{02}$ configurations and additional data on migration paths and charge density differences.

\bibitem{SCO_OV} H. A. Tahini, X. Tan, U. Schwingenschl\"ogl, and S. C. Smith, Formation and Migration of Oxygen Vacancies in SrCoO$_3$ and Their Effect on Oxygen Evolution Reactions, ACS Catal. 6, 5565 (2016)

\bibitem{O2CorrectionCeder} L. Wang, T. Maxisch and G. Ceder, Oxidation energies of transition metal oxides within the GGA+U framework, Phys. Rev. B 73, 195107 (2006)

\bibitem{O2Correction2} Y.-L. Lee, J. Kleis, J. Rossmeisl and D. Morgan, \textit{Ab initio} energetics of LaBO$_3$ (001) (B = Mn, Fe, Co, and Ni) for solid oxide fuel cell cathodes, Phys. Rev. B 80, 224101 (2009)

\bibitem{eformationdata} Emery, A. A. \& Wolverton, C. High-throughput DFT calculations of formation energy, stability and oxygen vacancy formation energy of ABO$_3$ perovskites, Sci Data 4, 170153 (2017)
  
\bibitem{phases} S. Xu, R. Jacobs, and D. Morgan, Factors Controlling Oxygen Interstitial Diffusion in the Ruddlesden-Popper Oxide La$_{2-x}$Sr$_x$NiO$_{4+\delta}$, Chem. Mater. 30, 7166 (2018)

\bibitem{RestaVanderbiltBook} R. Resta and D. Vanderbilt,  Theory of Polarization: A Modern Approach. In: Physics of Ferroelectrics. Topics in Applied Physics, vol 105. Springer (2007)

\bibitem{VanderbiltBook}  D. Vanderbilt, Berry Phases in Electronic Structure Theory: Electric Polarization, Orbital Magnetization and Topological Insulators. Cambridge: Cambridge University Press (2018)

\bibitem{Pol2Dsym}  P. Jadaun, D. Xiao, Q. Niu and S. K. Banerjee, Topological classification of crystalline insulators with space group symmetry, Phys. Rev. B 88, 085110 (2013)

\bibitem{Persistence_vac_20017} A. Raeliarijaona and  H. Fu, Persistence of strong and switchable ferroelectricity despite vacancies, Sci. Rep. 7, 41301 (2017)

\bibitem{STO111dft} S. E. Reyes-Lillo,  K. M. Rabe, and J. B. Neaton, Ferroelectricity in [111]-oriented epitaxially strained SrTiO$_3$ from first principles, Phys. Rev. Mater. 3, 030601 (2019)

\bibitem{NEB_LSFO} A. M. Ritzmann, A. B. Mu\~noz-Garc\'ia, M. Pavone, J. A. Keith and E. A. Carter, Ab Initio DFT+U Analysis of Oxygen Vacancy Formation and Migration in La$_{1-x}$Sr$_{x}$FeO$_{3-\delta}$ ($x = 0$, 0.25, 0.50), Chem. Mater. 25, 3011 (2013)

\bibitem{NEB_polaron} H. A. Tahini, X. Tan, S. N. Lou, J. Scott, R. Amal, Y. H. Ng and S. C. Smith, Mobile Polaronic States in $\alpha-$MoO$_3$: An ab Initio Investigation of the Role of Oxygen Vacancies and Alkali Ions, ACS Appl. Mater. Interfaces 8, 10911 (2016) 

\bibitem{He:2015dp} X. He and Y. Mo, Accelerated materials design of Na$_{0.5}$Bi$_{0.5}$TiO$_{3}$ oxygen ionic conductors based on first principles calculations, Phys. Chem. Chem. Phys. 17, 18035 (2015).

\bibitem{Anonymous:2016eh} T. T. Mayeshiba, D. D. Morgan, Factors controlling oxygen migration barriers in perovskites, Solid State Ion. 296, 71 (2016).

\bibitem{Chroneos:im} A. Chroneos, R.V. Vovk, I.L. Goulatis, and L.I. Goulatis, Oxygen transport in perovskite and related oxides: A brief review, J. Alloy. Compd. 494, 190 (2010)

\bibitem{deSouza:2015eb} R. A De Souza, Oxygen Diffusion in SrTiO$_3$ and Related Perovskite Oxides, Adv. Funct. Mater. 25, 6326 (2015)

\bibitem{DeSouza:2012jl} R. A. De Souza, A. Ramadan and S. H\"orner,  Modifying the barriers for oxygen-vacancy migration in fluorite-structured CeO$_2$ electrolytes through strain: a computer simulation study, Energy Environ. Sci. 5, 5445 (2012).

\bibitem{Vives:2015dd} S. Vives and C. Meunier, Defect cluster arrangements and oxygen vacancy migration in Gd doped ceria for different interatomic potentials, Solid State Ion. 283, 137 (2015).

\bibitem{Kong:2019gm} F. Kong, C. Liang, L. Wang, Y. Zheng, S. Perananthan, R. C. Longo, J. P. Ferraris, M. Kim and K. Cho, Kinetic Stability of Bulk LiNiO$_2$ and Surface Degradation by Oxygen Evolution in LiNiO$_2$-Based Cathode Materials, Adv. Energy Mater. 9, 1802586 (2019).

\bibitem{Lu:eb} W. L\"u, C. Li, L. Zheng, J. Xiao, W. Lin, Q. Li, Xiao Renshaw Wang, Z. Huang, S. Zeng, K. Han, W. Zhou, K. Zeng, J. Chen, Ariando, W. Cao and T. Venkatesan, Multi‐Nonvolatile State Resistive Switching Arising from Ferroelectricity and Oxygen Vacancy Migration, Adv. Mater. 29, 1606165 (2017)

\bibitem{Liao:ki} Z. Liao, P. Gao, X. Bai, D. Chen and J. Zhang, Evidence for electric-field-driven migration and diffusion of oxygen vacancies in Pr$_{0.7}$Ca$_{0.3}$MnO$_{3}$, J. Appl. Phys. 111, 114506 (2012)

\bibitem{Hanzig:ch} J. Hanzig, E. Mehner, S. Jachalke, F. Hanzig, M. Zschornak, C. Richter, T. Leisegang, H. St\"ocker and D. C Meyer, Dielectric to pyroelectric phase transition induced by defect migration, New J. Phys 17, 023036 (2015)

\bibitem{Leighton:2019bo} C. Leighton, Electrolyte-based ionic control of functional oxides, Nat. Mater. 18, 13 (2019).

\bibitem{Lee:2018dr} J. Lee, E. Ahn, Y.-S. Seo, Y. Kim, T.-Y. Jeon, J. Cho, I. Lee and H. Jeen, Redox-Driven Nanoscale Topotactic Transformations in Epitaxial SrFe$_{0.8}$Co$_{0.2}$O$_{3-x}$ under Atmospheric Pressure, Phys. Rev. Applied, 10, 054035 (2018). Phys. Rev. Applied 11, 059901(E) (2019)

\bibitem{STC_ovac} C. Mitra, C. Lin, A. B. Posadas, and A. A. Demkov, Role of Oxygen Vacancies in Room-Temperature Ferromagnetism in Cobalt-Substituted SrTiO$_3$, Phys. Rev. B 90, 125130 (2014).

\bibitem{STC_ovac_exp} A. B. Posadas, C. Mitra, C. Lin, A. Dhamdhere, D. J. Smith, M. Tsoi, and A. A. Demkov, Oxygen Vacancy-Mediated Room-Temperature Ferromagnetism in Insulating Cobalt-Substituted SrTiO$_3$ Epitaxially Integrated with Silicon, Phys. Rev. B 87, 144422 (2013).

\bibitem{STO_qparael} D. Shin, S. Latini, C. Sch\"afer, S. A. Sato, U. De Giovannini, H. H\"ubener, and A. Rubio, Quantum Paraelectric Phase of SrTiO$_3$ from First Principles, Phys. Rev. B 104, L060103 (2021).

\bibitem{STO_photogs} S. Latini, D. Shin, S. A. Sato, C. Sch\"afer, U. De Giovannini, H. H\"ubener, and A. Rubio, The Ferroelectric Photo Ground State of SrTiO$_3$: Cavity Materials Engineering, Proc Natl Acad Sci USA 118, e2105618118 (2021).

\bibitem{STO_mfdefects} T. Xu, T. Shimada, M. Mori, G. Fujimoto, J. Wang, and T. Kitamura, Defect Engineering for Nontrivial Multiferroic Orders in SrTiO$_3$, Phys. Rev. Materials 4, 124405 (2020).

\bibitem{metallicferro1} W. X. Zhou and A. Ariando 2020, Review on ferroelectric/polar metals, Jpn. J. Appl. Phys. 59 SI0802.
 
\bibitem{metallicferro2} Hadjimichael, M. et al. Metal–ferroelectric supercrystals with periodically curved metallic layers. Nat Mater 20, 495–502 (2021).

\bibitem{ordercatios-1} Inkinen, S., Yao, L. \& Dijken, S. van. Reversible thermal strain control of oxygen vacancy ordering in an epitaxial ${\rm{L}}{{\rm{a}}_{0.5}}{\rm{S}}{{\rm{r}}_{0.5}}{\rm{Co}}{{\rm{O}}_{3 - \delta }}$ film. 1–8 (2020).
  
\bibitem{ordercatios-2} Ning, Q. Z. C. O. R. C. X. Z. and C. A. R. S. Voltage Control of Magnetism above Room Temperature in Epitaxial SrCo. 1–9 (2020).

\bibitem{JT} Jahn H. A.  and Teller E.  1937 Stability of polyatomic molecules in degenerate electronic states - I—Orbital degeneracyProc. R. Soc. Lond. A161220–235

\bibitem{CBFerro} Katarzyna Tkacz-Śmiech, A. Koleżyński \& W. S. Ptak (2000) Chemical bond in ferroelectric perovskites, Ferroelectrics, 237:1, 57-64

\bibitem{breakingoxydifucion2021} Kanishk Rawat, Dillon D. Fong, and Dilpuneet S. Aidhy , "Breaking atomic-level ordering via biaxial strain in functional oxides: A DFT study", Journal of Applied Physics 129, 095301 (2021)

\bibitem{chemicalorderin2021nature} Iijima, S., Yang, W., Matsumura, S. et al. Atomic resolution imaging of cation ordering in niobium–tungsten complex oxides. Commun Mater 2, 24 (2021). 

\bibitem{hoppingconductivity} Jukichi Hombo, Yasumichi Matsumoto, Takeo Kawano, Electrical conductivities of SrFeO$_{3-\delta}$ and BaFeO$_{3-\delta}$ perovskites, Journal of Solid State Chemistry, Volume 84, Issue 1,1990,

\bibitem{otroferrovalue} Zhang, B. H., Liu, X. Q. \& Chen, X. M. Review of experimental progress of hybrid improper ferroelectricity in layered perovskite oxides. J Phys D Appl Phys 55, (2022).
  

\end{thebibliography}
\end{document}


\title{Supplemental Material:    Oxygen deficiency and migration mediated electric polarization in Fe,Co-substituted SrTiO$_{3-\delta}$}
\author{Emilio A. Cort\'es Estay}
\affiliation{Departamento de F\'isica, Universidad T\'ecnica Federico Santa Mar\'ia, Espa\~na 1680, Valpara\'iso, P.O. Box 110-V, Chile}
\author{Shyue Ping Ong}
\affiliation{Department of NanoEngineering, University of California, San Diego, 9500 Gilman Drive, La Jolla, California 92093, USA}
\author{Caroline. A. Ross}
\affiliation{Department of Materials Science and Engineering, Massachusetts Institute of Technology, 77 Massachusetts Avenue, Cambridge, Massachusetts 02139, USA}
\author{Juan M. Florez}
\email{juanmanuel.florez@usm.cl, jmflorez@mit.edu}
\affiliation{Departamento de F\'isica, Universidad T\'ecnica Federico Santa Mar\'ia, Espa\~na 1680, Valpara\'iso, P.O. Box 110-V, Chile}
\affiliation{Department of Materials Science and Engineering, Massachusetts Institute of Technology, 77 Massachusetts Avenue, Cambridge, Massachusetts 02139, USA}

\date{\today}
\pacs{}
\maketitle 

\onecolumngrid

\clearpage
\section{Configurations of oxygen vacancies for $\delta_{02}$}
\begin{figure*}[h!] 
\vspace{15pt}
\centering
\includegraphics[width=0.9\linewidth, trim = 1.5cm 1cm 1cm 2.8cm, clip]{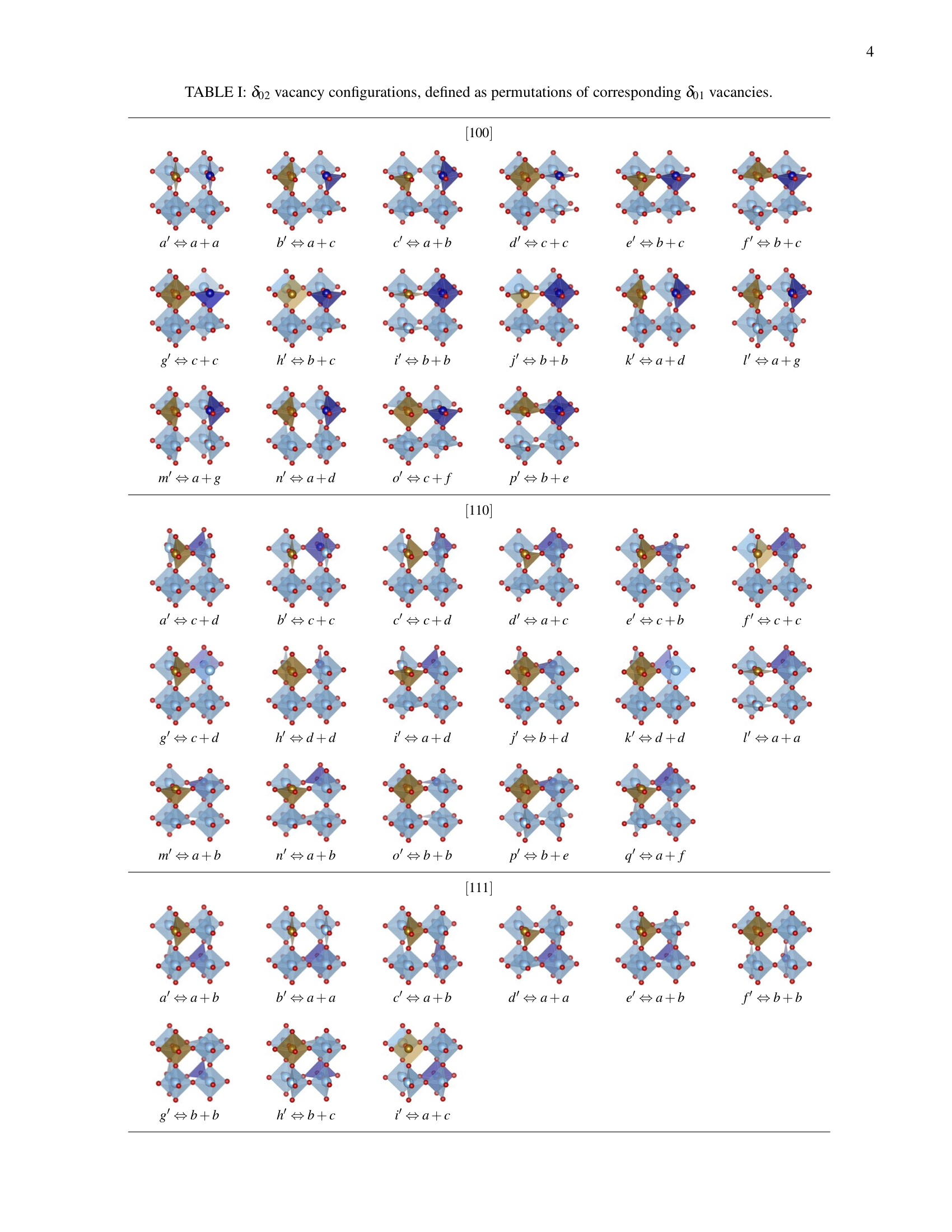}
\caption{Configurations of oxygen vacancies for $\delta_{02}$ systems in main text Figure 1. For each Fe-Co orientation considered in this work, we labeled the different $\delta_{02}$ combinations in terms of the $\delta_{01}$ vacancies displayed in Figure 1.}
\end{figure*}

\clearpage
\section{GGA+U: Selection of Hubbard $+U$ parameters}
\begin{figure*}[h!] 
\vspace{15pt}
\centering
\includegraphics[width=0.9\linewidth]{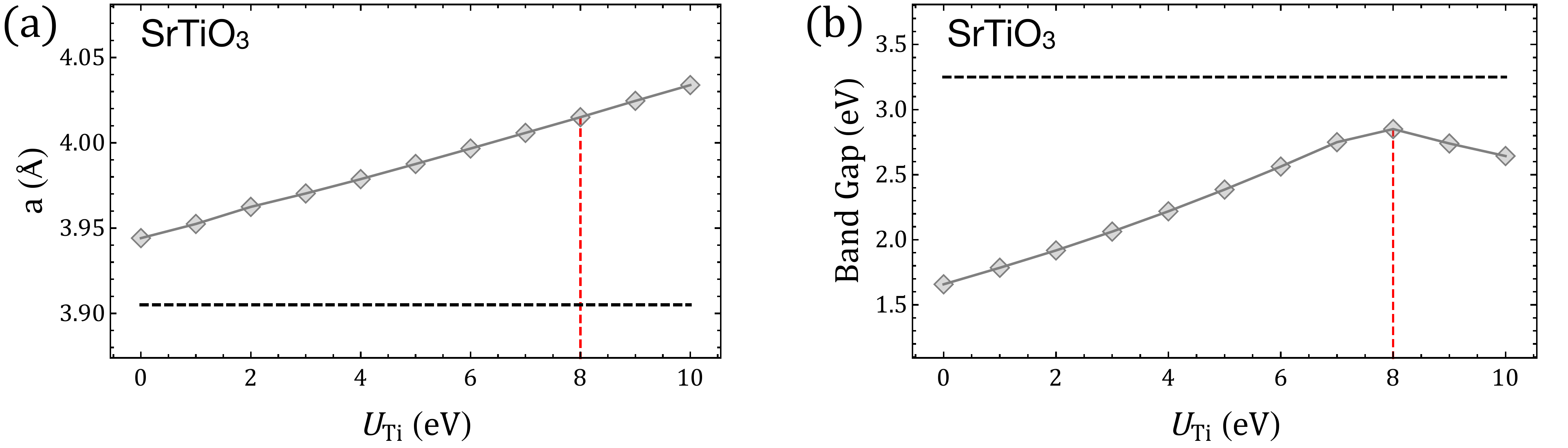}
\includegraphics[width=0.9\linewidth]{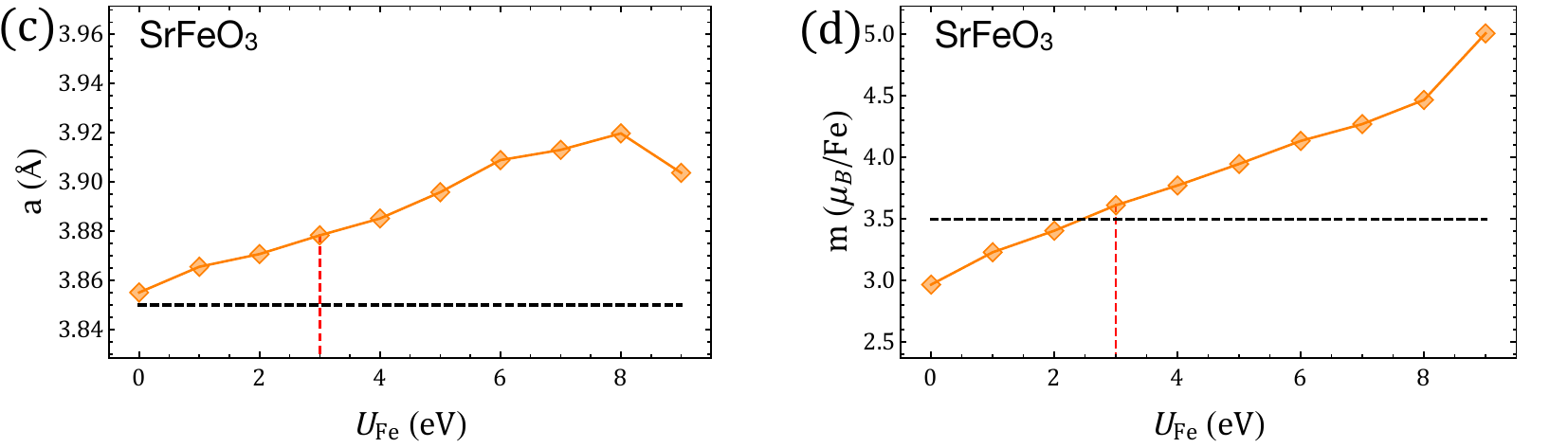}
\includegraphics[width=0.9\linewidth]{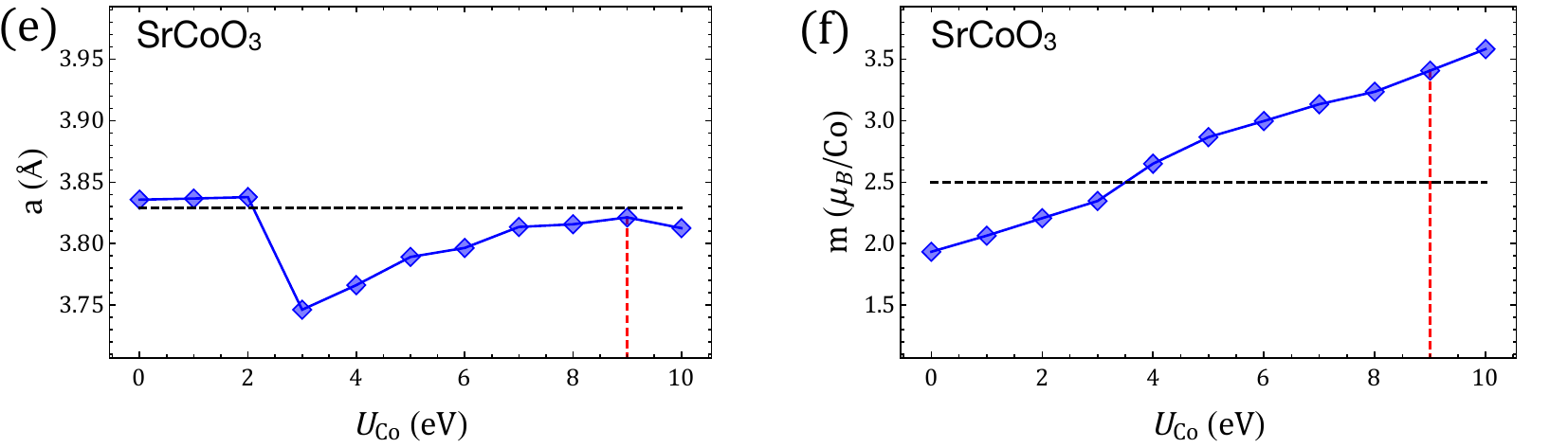}
\caption{Properties of SrTiO$_{3}$, SrFeO$_{3}$ and SrCoO$_{3}$ perovskites calculated with GGA+U using different Hubbard paramterers for Ti, Fe and Co. (a,b) Lattice parameter and band gap of SrTiO$_3$. (c-f) Lattice parameter and magnetic moment per Fe/Co ion of SrFeO$_3$ and SrCoO$_3$, respectively. Horizontal lines indicate the respective experimental values (from Refs. for SrTiO$_{3}$\cite{STO_Review}, SrFeO$_{3}$\cite{ishiwata_versatile_2011,kinoshita_contrasting_2016}, SrCoO$_{3}$\cite{long_synthesis_2011}) while vertical lines indicate the $U$ values chosen for this work.}
\end{figure*}

\begin{figure*}[h!] 
	\centering
	\includegraphics[width=0.9\linewidth]{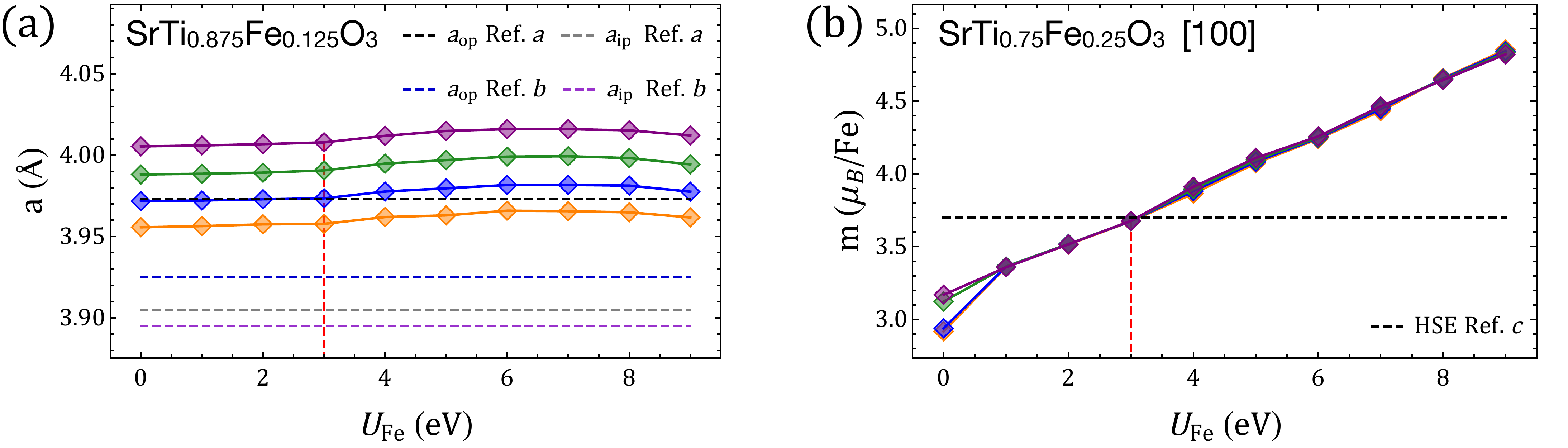}
	\caption{Properties of SrTi$_{1-x}$Fe$_{x}$O$_{3}$ perovskites ($x=\{0.125, 0.25\}$) calculated with GGA+U using different Hubbard paramterers for Ti and Fe. (a) Lattice parameter of SrTi$_{0.875}$Fe$_{0.125}$O$_{3}$ compared with experimental parameters for $x=0.13$ (from Refs. $a$\cite{roomt1} and $b$\cite{STF_molecular}). (b) Magnetic moment per Fe ion for $x=0.25$ compared with HSE results (Ref. $c$ \cite{OdefSTF}).}
\end{figure*}

\begin{figure*}[h!] 
	\centering
	\includegraphics[width=0.9\linewidth]{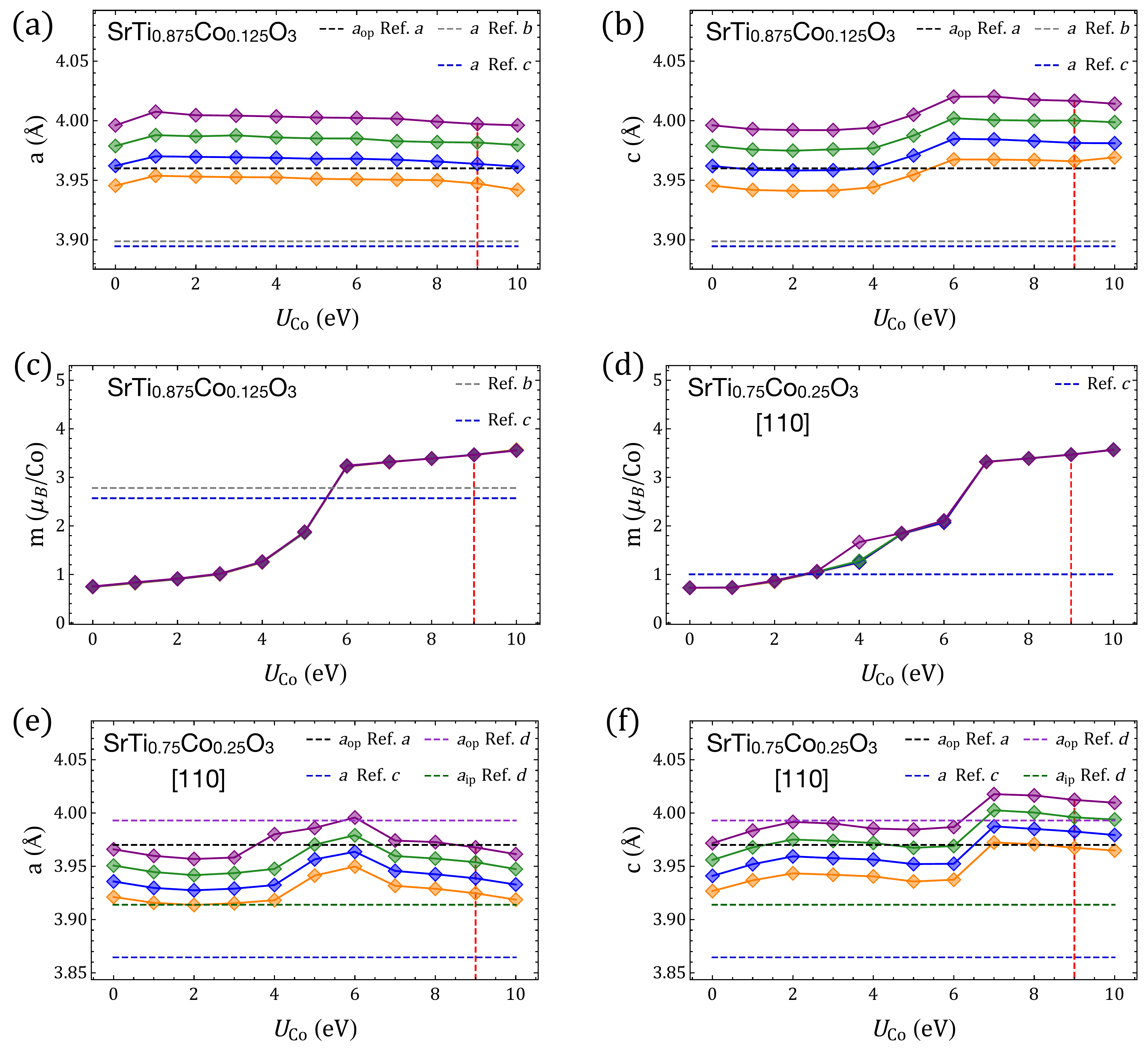}
	\caption{Properties of SrTi$_{1-x}$Co$_{x}$O$_{3}$ perovskites ($x=\{0.125, 0.25\}$) calculated with GGA+U using different Hubbard paramterers for Ti and Co. (a,b) Lattice parameters of SrTi$_{0.875}$Co$_{0.125}$O$_{3}$ compared with experimental parameters for $x=0.14$ (from Refs. $a$\cite{bi_structure_2010} and $b$\cite{pascanut_magnetic_2006}) and HSE results (Ref. $c$\cite{STC} ). (c,d) Magnetic moment per Co ion for $x=0.125$ and $x=0.25$. (e,f) Lattice parameters for $x=0.25$ and $[110]$ aligned Co-Co, compared with experimental values for $x=0.23$ (Refs. $a$ and $d$\cite{roomt1}) and HSE calculations for $x=0.25$ (Ref. $c$).}
\end{figure*}

\clearpage
\section{Oxygen-vacancy migration}
In this section we present complementary information associated with the oxygen migrations shown in the Figure 2 of the article main text as well as the results for two additional migration paths for $\delta_{01}$ systems. In all the cases the intermediate structural images representing the $v_\mathrm{O}$ migration are obtained through: (i) a linear interpolation, between the end-structures, of the atomic coordinates and lattice vectors; (ii) a relaxation of such interpolation using the Nudged Elastic Band method (NEB). Depending on the end-states magnetic ordering for $\delta_{01}$ and $\delta_{02}$ respectively, which are detailed in Table I of the article, the migration images were annotated FM or AFM. 

\subsection{Migration path between $\delta_{01}$ vacancies in Figure 2: $[100]_{a}$ and $[100]_{c}$}

\begin{figure*}[!htb]
	\centering
	\begin{minipage}{1.0\linewidth}
	\includegraphics[width=1.0\linewidth]{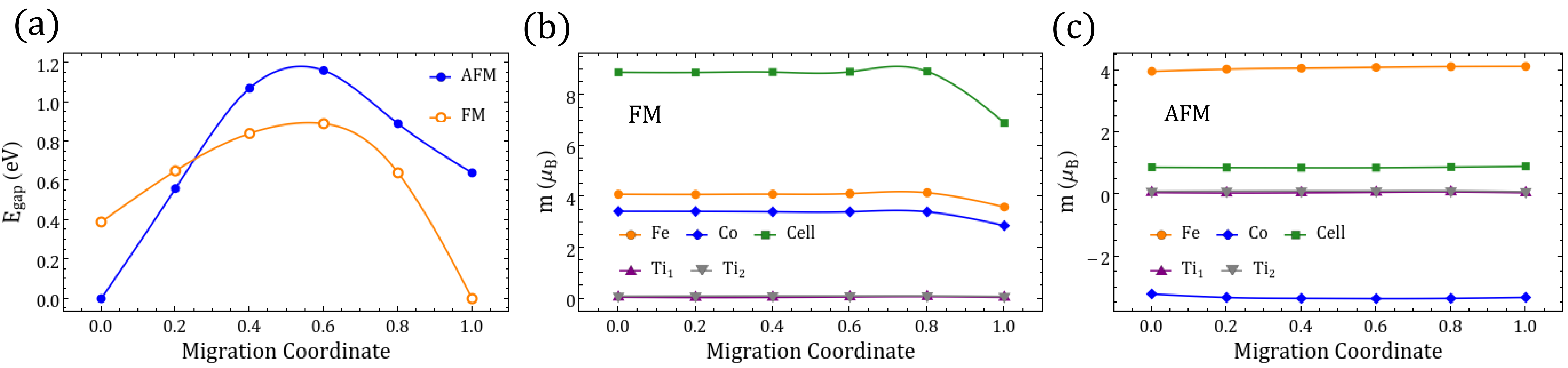}
	\end{minipage}
	\caption{(a) Band gap $E_{gap}$ along the migration path between $[100]_{a}$ (migration coordinate $0.0$) and $[100]_{c}$ (migration coordinate $1.0$) for FM and AFM NEB-relaxed structures. 
	(b,c) Magnetic moment ``m'' of the TM labeled in Figure 2 of the main text, and of the whole perovskite supercell for the relaxed FM and AFM migration paths. End-states values are associated with the gs reported in Table I.}
	\label{fig:1V_Mig_ac}
\end{figure*}
\subsection{Migration path between $\delta_{02}$ vacancies in Figure 2: $[100]_{b'}$ and $[100]_{e'}$}
\begin{figure*}[!htb]
	\centering
	\begin{minipage}{1.0\linewidth}
	\includegraphics[width=1.0\linewidth]{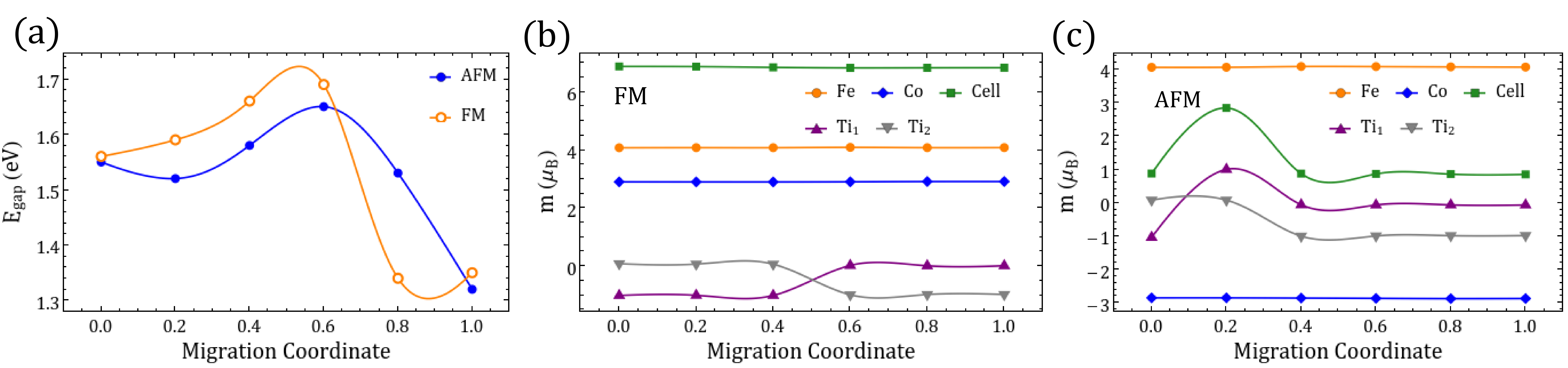}
	\end{minipage}
	\caption{(a) Band gap $E_{gap}$ along the migration path between $[100]_{b'}$ (migration coordinate $0.0$) and $[100]_{e'}$ (migration coordinate $1.0$) for FM and AFM NEB-relaxed structures. 
	(b,c) Magnetic moment ``m'' of the TM labeled in Figure 2 of the main text, and of the whole perovskite supercell for the relaxed FM and AFM migration paths. End-states values are associated with the gs reported in Table I.}
	\label{fig:2V_Mig_be}
\end{figure*}
\vspace{55pt}
\clearpage
\subsection{Other migration paths for $\delta_{01}$}
\subsubsection{Migration connecting symmetry-equivalent vacancies: the case of $[100]_{c}$}
\vspace{-20pt}
\begin{figure*}[!htb]
	\centering
	\includegraphics[width=1.0\linewidth]{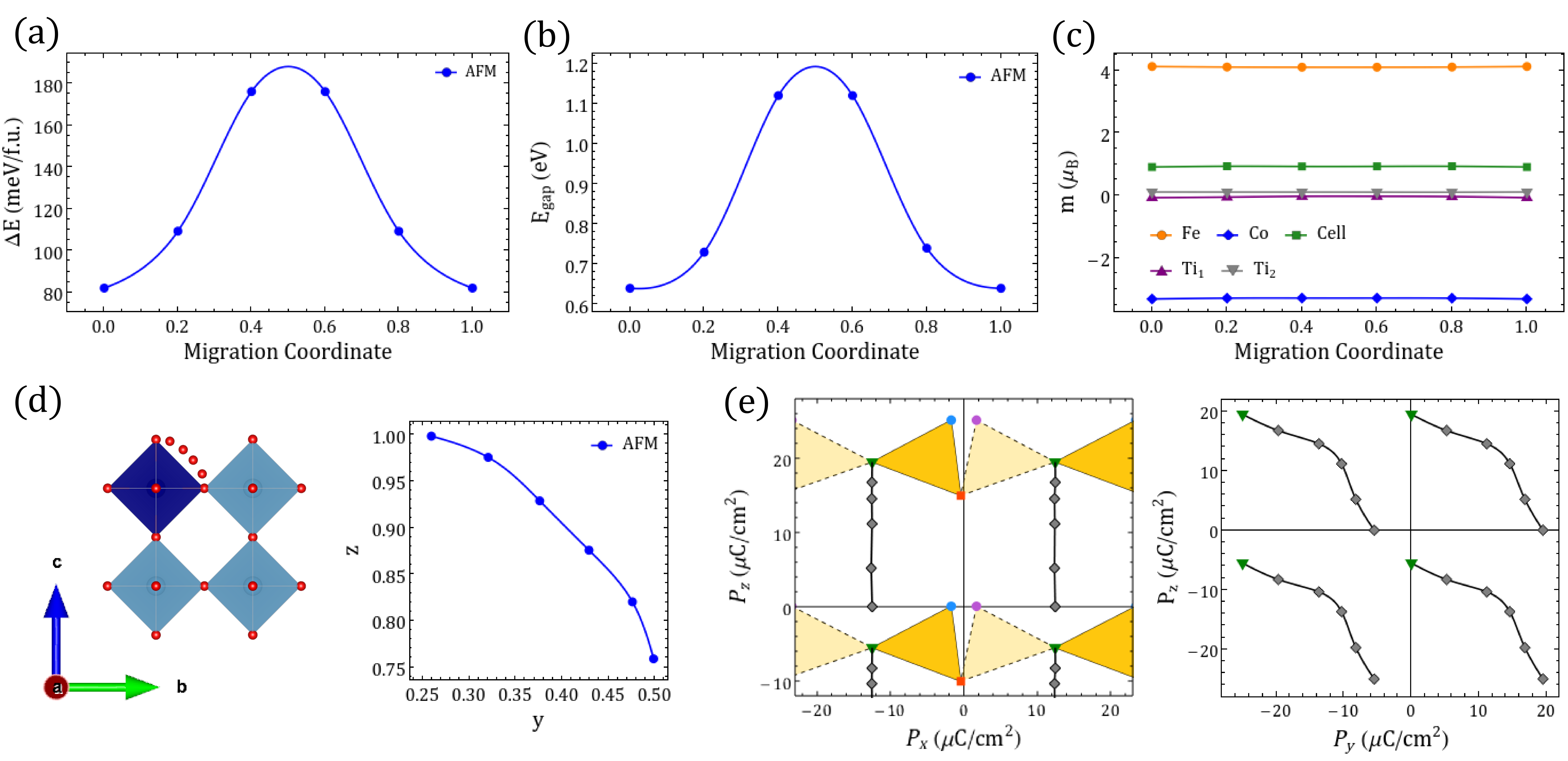}
	\vspace{-20pt}
	\caption{(a) Energy relative to the $\delta_{01}$ gs, (b) band gap $E_{gap}$ and (c) magnetic moments ``m'' along the migration path between $[100]_{c}$ (migration coordinate $0.0$) and a symmetrically equivalent structure (migration coordinate $1.0$). (d) Oxygen migration path in which Co-$v_\mathrm{O}$-Ti switches from the $c$ direction to the $b$ direction, and the corresponding relaxed fractional position of the migrated oxygen. 
	(e) Electric polarization lattices along NEB-relaxed paths.}
	\label{fig:1V_Mig_rot}
\end{figure*}

\vspace{-10pt}
\subsubsection{Migration connecting non-equivalent vacancies: the case of $[100]_{a}$ and $[100]_{b}$}
\vspace{-20pt}
\begin{figure*}[!htb]
	\centering
	\includegraphics[width=1.0\linewidth]{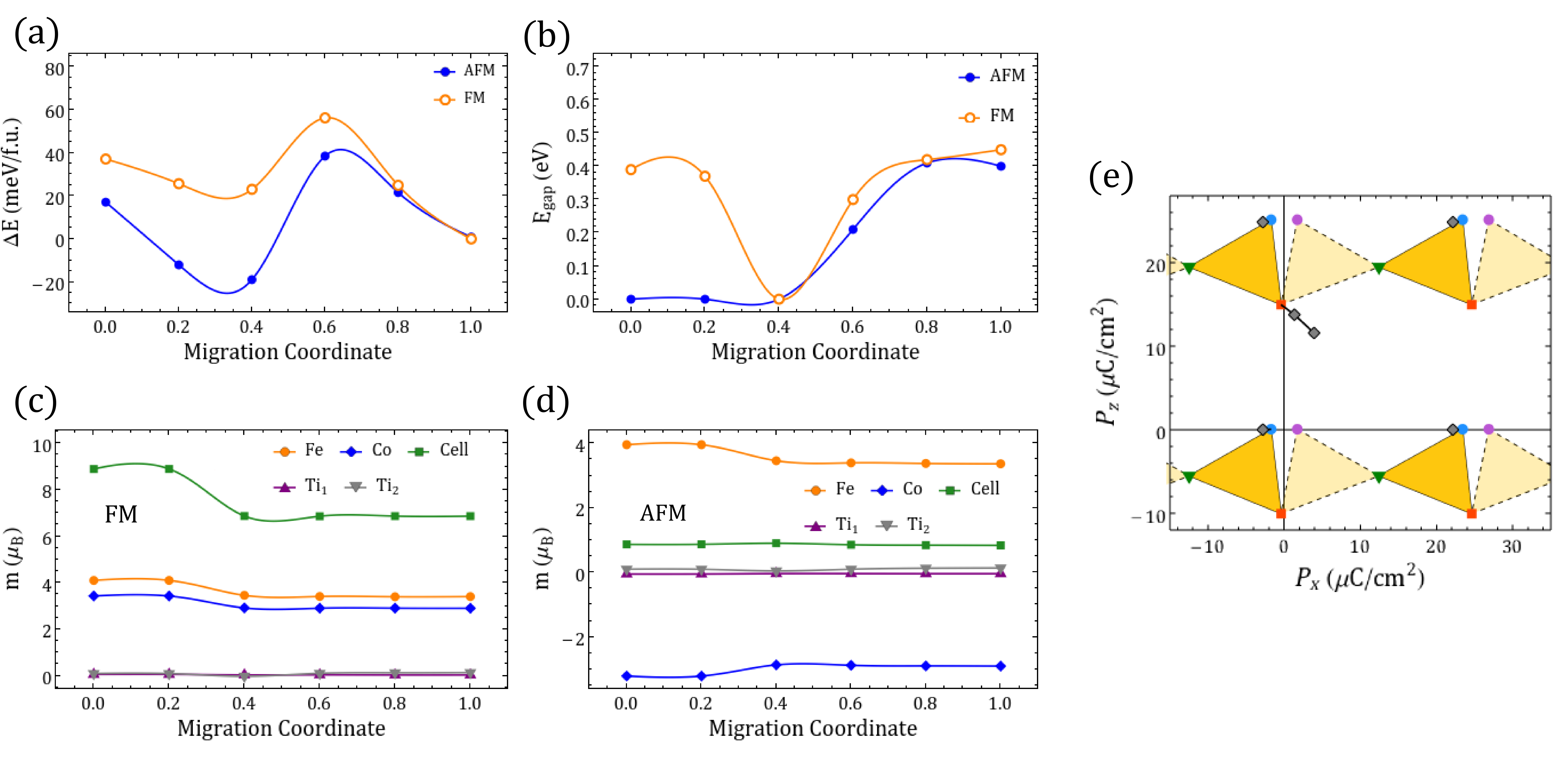}
	\vspace{-20pt}
	\caption{(a) Energy relative to the $\delta_{02}$ gs, (b) band gap $E_{gap}$ and (c,d) magnetic moments ``m'' along the migration path between $[100]_{a}$ (migration coordinate $0.0$) and $[100]_{b}$ (migration coordinate $1.0$), for AFM and FM NEB-relaxed structures. (e) Electric polarization lattices corresponding to the FM NEB-relaxed path. }
	\label{fig:1V_Mig_ab}
\end{figure*}

\clearpage
\subsection{Other migration paths for $\delta_{02}$}
\subsubsection{Migration path between $\delta_{02}$ vacancies: $[100]_{b'}$ and $[100]_{m'}$}
\vspace{-20pt}
\begin{figure*}[!htb]
	\centering
	\includegraphics[width=1.0\linewidth]{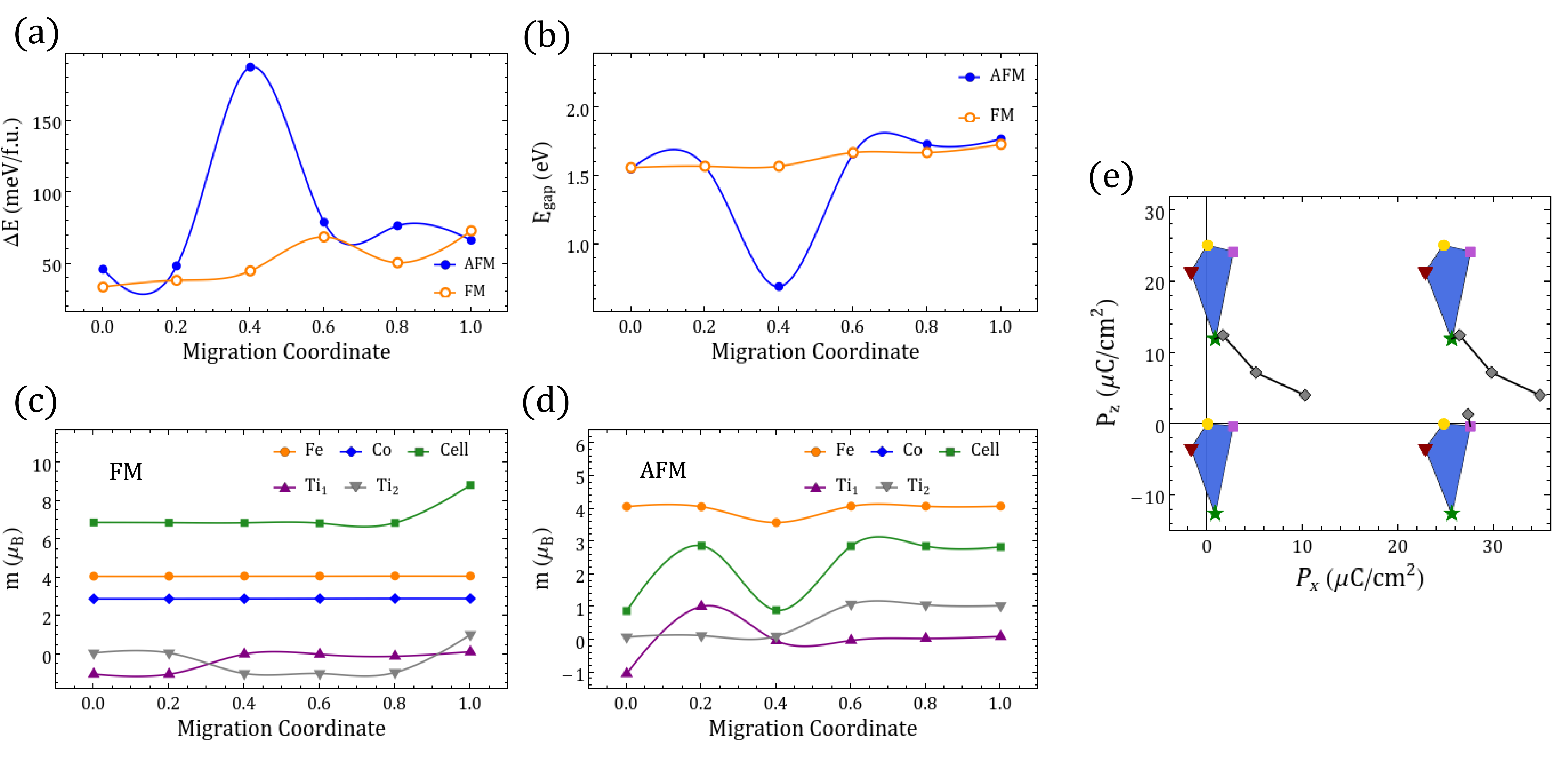}
	\vspace{-20pt}
	\caption{(a) Energy relative to the $\delta_{02}$ gs, (b) band gap $E_{gap}$ and (c,d) magnetic moments ``m'' along the migration path between $[100]_{b'}$ (migration coordinate $0.0$) and $[100]_{m'}$ (migration coordinate $1.0$), for AFM and FM NEB-relaxed structures. (e) Electric polarization lattices corresponding to the FM NEB-relaxed path. }
	\label{fig:2V_Mig_bm}
\end{figure*}

\vspace{-10pt}
\subsubsection{Migration path between $\delta_{02}$ vacancies: $[100]_{a'}$ and $[100]_{b'}$}
\vspace{-20pt}
\begin{figure*}[!htb]
	\centering
	\includegraphics[width=1.0\linewidth]{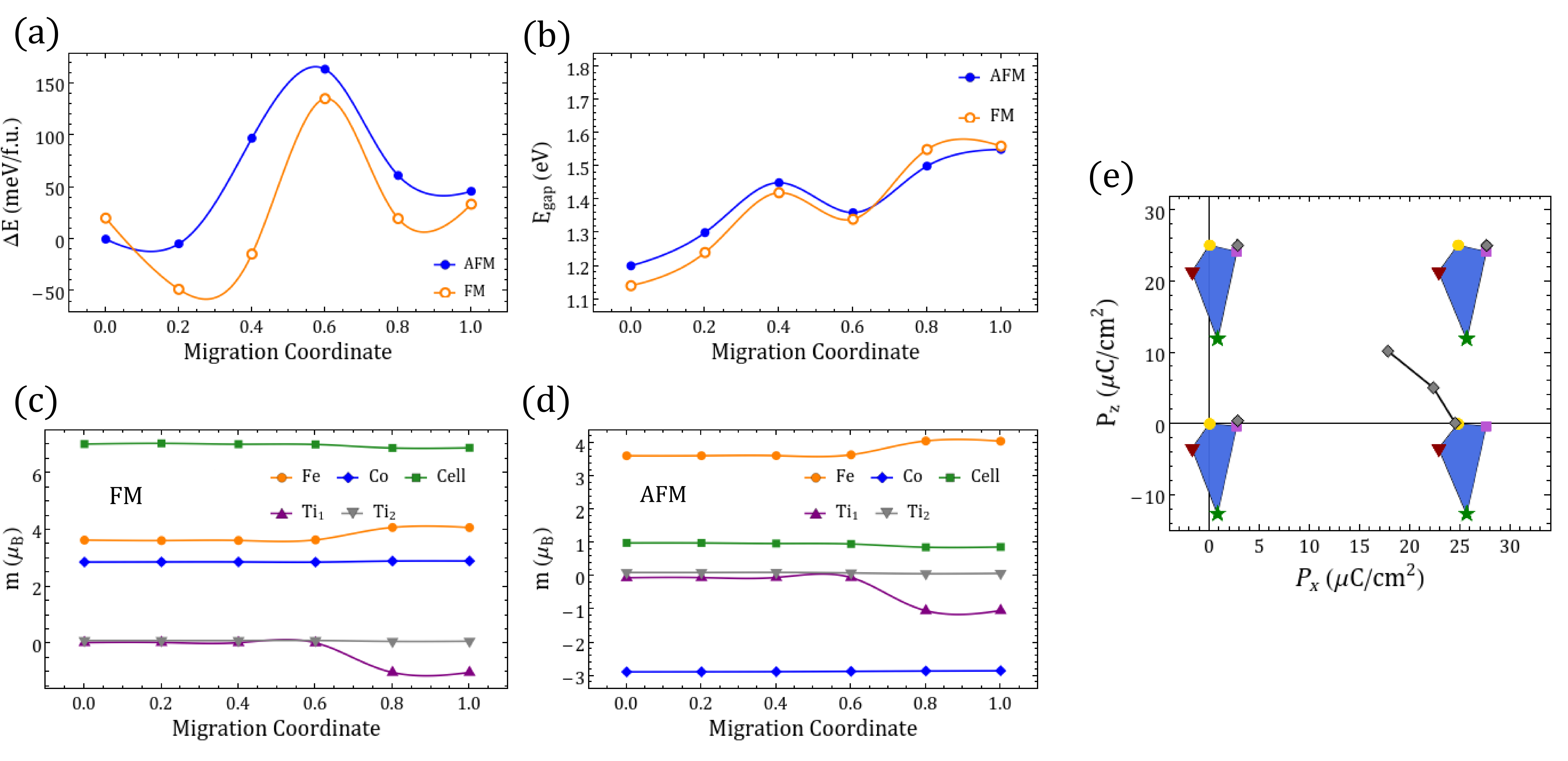}
	\vspace{-20pt}
	\caption{(a) Energy relative to the $\delta_{02}$ gs, (b) band gap $E_{gap}$ and (c,d) magnetic moments ``m'' along the migration path between $[100]_{a'}$ (migration coordinate $0.0$) and $[100]_{b'}$ (migration coordinate $1.0$), for AFM and FM NEB-relaxed structures. (e) Electric polarization lattices corresponding to the FM NEB-relaxed path. }
	\label{fig:2V_Mig_ab}
\end{figure*}

\clearpage
\section{Projected Density of States}
\begin{figure*}[!htb]
	\centering
	\includegraphics[width=0.48\linewidth]{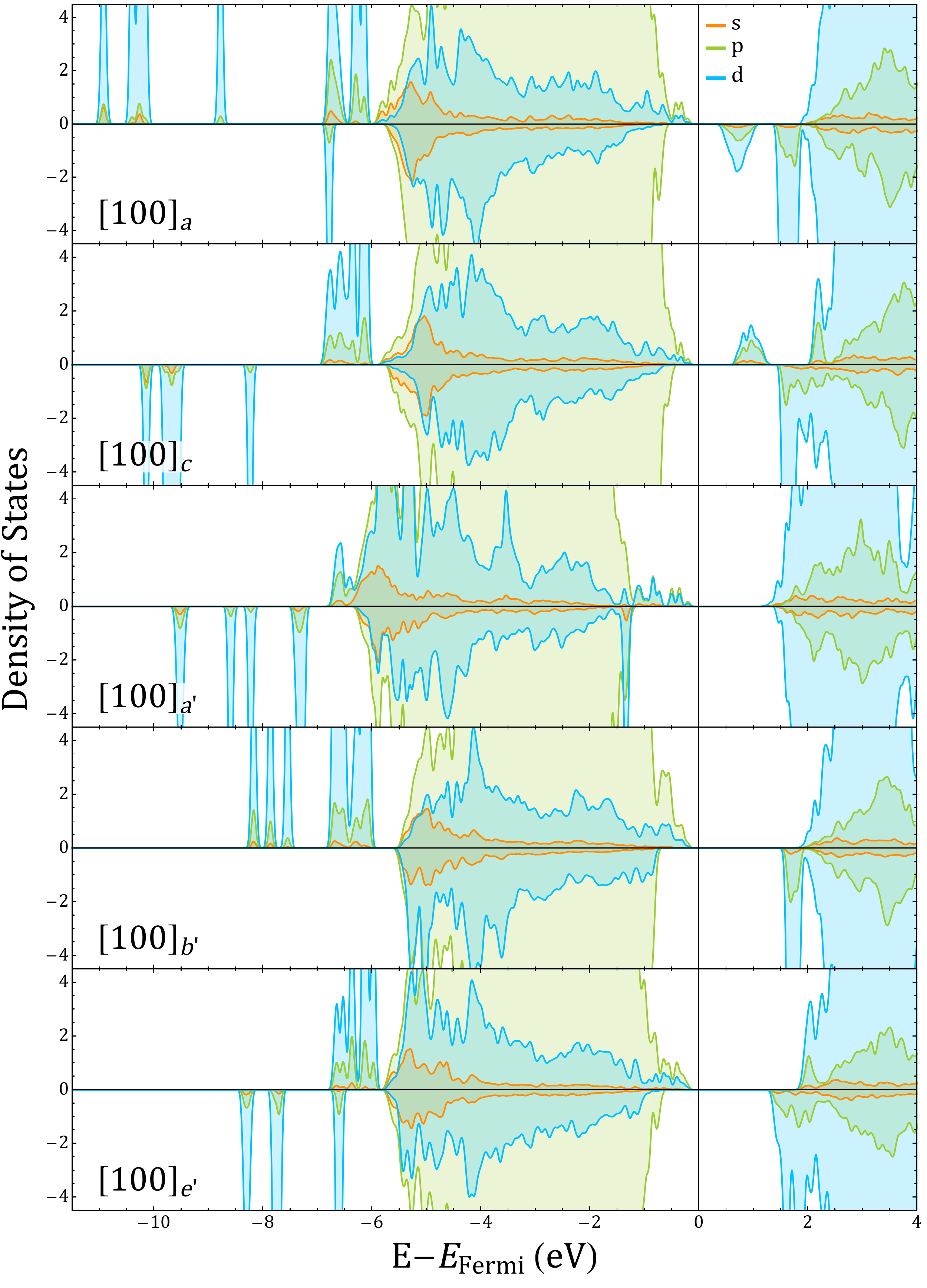}
	\hspace{10pt}
	\includegraphics[width=0.48\linewidth]{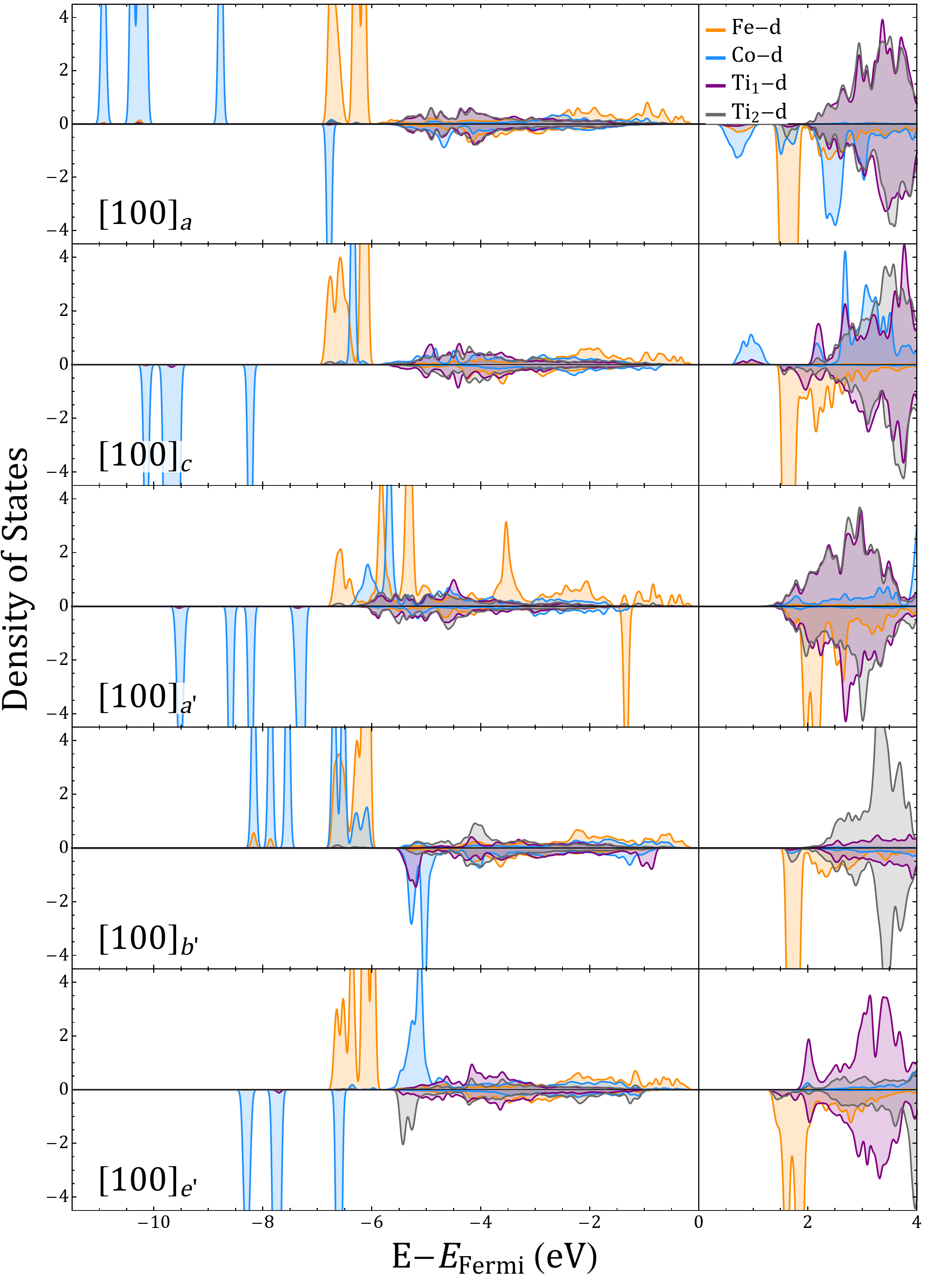}
	\caption{$spd$-decomposed density of states (left panel) and Fe, Co, Ti$_1$ and Ti$_2$ $d$-orbital projected density of states of (right panel) for $[100]_{a,c}$ and $[100]_{a',b',e'}$
	vacancies corresponding to $\delta_{01}$ and $\delta_{02}$, respectively. }
	\label{fig:projected_dos}
\end{figure*}